\documentclass[12pt,reqno]{amsart}
\usepackage{amsmath,amssymb,amsfonts,amsthm}
\usepackage[mathscr]{eucal}
\usepackage[all]{xy}
\usepackage{hyperref}
\usepackage{setspace}
\usepackage{bm}
\usepackage{xcolor}
\allowdisplaybreaks

\author{V.A.Abakumova, S.L.~Lyakhovich}

\address{Physics Faculty, Tomsk State University, Lenin ave. 36, Tomsk 634050, Russia.}

\email{abakumova@phys.tsu.ru, \, sll@phys.tsu.ru}

\title{Hamiltonian constraints and unfree gauge symmetry}

\begin{document}

\maketitle

\begin{abstract}
We study Hamiltonian form of unfree gauge symmetry  where the gauge
parameters have to obey differential equations. We consider the
general case such that the Dirac-Bergmann algorithm does not
necessarily terminate at secondary constraints, and tertiary and
higher order constraints may arise. Given the involution relations
for the first-class constraints of all generations, we provide
explicit formulas for unfree gauge transformations in the
Hamiltonian form, including the differential equations constraining
gauge parameters. All the field theories with unfree gauge symmetry
share the common feature: they admit sort of ``global constants of
motion'' such that do not depend on the local degrees of freedom.
The simplest example is the cosmological constant in the unimodular
gravity. We consider these constants as modular parameters rather
than conserved quantities. We provide a systematic way of
identifying all the modular parameters. We demonstrate that the
modular parameters contribute to the Hamiltonian constraints, while
they are not explicitly involved in the action. The Hamiltonian
analysis of the unfree gauge symmetry is precessed by a brief
exposition for the Lagrangian analogue, including explicitly
covariant formula for degrees of freedom number count. We also adjust
the BFV-BRST Hamiltonian quantization method for the case of unfree
gauge symmetry. The main distinction is in the content of the
non-minimal sector and gauge fixing procedure. The general formalism
is exemplified by traceless tensor fields of irreducible spin $s$
with the gauge symmetry parameters obeying transversality equations.
\end{abstract}
\section{Introduction}
\noindent Gauge symmetry is usually understood as a set of the
infinitesimal transformations of the fields such that leaves the
action intact, while the transformation parameters are the functions
of space-time. Gauge symmetry is said unfree if the invariance of
the action requires the gauge parameters to obey the PDE system. The
general solution of the equations constraining gauge parameters
must involve arbitrary functions of all $d$ space-time coordinates.
If the solution includes arbitrary functions of $d-1$ coordinates or
less, then this is not gauge symmetry.

The most known example of an unfree  gauge symmetry is the
volume-preserving diffeomorphism of unimodular gravity (UG). Various
analogues of the linearized UG \cite{Alvarez:2006uu},
\cite{Blas:2007pp} are known among the free higher spin field
theories, with gauge parameters constrained by transversality
equations \cite{SKVORTSOV2008301}, \cite{Campoleoni2013}. The key
distinction of UG from General Relativity (GR) with $\Lambda$-term
is that $\Lambda$ is a specific constant fixed from the outset in
the action of GR, while UG comprises dynamics with any cosmological
constant. For discussion of the role of cosmological constant in UG
and further references, we cite \cite{Percacci2018}. Also
modifications of UG can be found in \cite{BARVINSKY201759},
\cite{PhysRevD.100.023542}, where $\Lambda$ is defined dynamically,
not as pre-fixed parameter in the action. All the field theories
with unfree gauge symmetry share the common feature: they admit the
``global constants of motion'' such that do not depend on the local
degrees of freedom, with $\Lambda$ of the UG being the simplest
example. This general fact is explained from various viewpoints in
the recent articles \cite{KAPARULIN2019114735},
\cite{Kaparulin2019}, \cite{Abakumova:2019uoo}. As the specific
values of these integration constants are defined by the field
asymptotics, not the Cauchy data, we consider them as modular
parameters rather than conserved quantities. In the higher spin
field analogues of UG, for example, similar modular parameters
exist, and their number grows with spin, although this fact has not
previously been noticed.

While the examples of unfree gauge symmetry have been known for a
long time, the general theory of this class of gauge systems began
to develop relatively recently. In the article
\cite{KAPARULIN2019114735},  general structure is established for
unfree gauge symmetry algebra in Lagrangian formalism, and the
modification is proposed for the Faddeev-Popov (FP) method such that
accounts for the constraints imposed on gauge parameters. In the
article \cite{Kaparulin2019}, the
BV-BRST\footnote{Batalin-Vilkovisky--Becchi-Rouet-Stora-Tyutin.}
field-antifield formalism is worked out for the systems with unfree
gauge symmetry. In the article \cite{Abakumova:2019uoo},  general
structures are identified in the algebra of Hamiltonian constraints
such that describe unfree gauge symmetry. Before this work, the
equations constraining gauge parameters in Hamiltonian formalism
have been unknown even in specific models. The article
\cite{Abakumova:2019uoo} assumes that the Dirac-Bergmann algorithm
terminates at secondary constraints, no tertiary ones are allowed.
In this work, we provide the Hamiltonian description of unfree gauge
symmetry in the general case, with the sequence of constraints of
any finite order. Besides the reason of generality, this is also
motivated by specific models. While in UG the Dirac-Bergmann
algorithm terminates at the stage of secondary constraints, in the
higher spin field theories with unfree gauge symmetry, the sequence
of constraints turns out linearly growing with spin, so the tertiary
constraints arise for $s=3$. The number of modular parameters is
also growing with spin, and they all contribute to the constraints.
The new phenomenon here is that the modular parameters, being
connected to the non-trivial asymptotics of the fields, can make the
constraints explicitly depending on the space-time point $x$, even
though the original Lagrangian is $x$-independent. This phenomenon
has previously unnoticed analogue in Lagrangian formalism.

The main goal of this article is to work out Hamiltonian description
of general unfree gauge symmetry. Then, we also extend the
BFV--BRST\footnote{Batalin-Fradkin-Vilkovisky--Becchi-Rouet-Stora-Tyutin.}
formalism to this class of theories, with main modifications related
to the non-minimal sector of ghosts. The general formalism is
exemplified by the massless spin-$s$ theory where the irreducible
representation is realized by traceless tensors
\cite{SKVORTSOV2008301}. To make the article self-contained we
precede the Hamiltonian description of unfree gauge symmetry with
the corresponding Lagrangian formalism mostly providing the facts
from \cite{KAPARULIN2019114735}, \cite{Kaparulin2019}, with a more
emphasis on modular parameters. We also provide a convenient formula
for the degree of freedom counting in Lagrangian formalism in the
case of unfree gauge symmetry.

\section{Unfree gauge symmetry in Lagrangian formalism: \\
completion functions, and modular parameters}

\noindent Unfree gauge symmetry is a deviation from the usual
assumptions implied by general theory of gauge systems as it is
formulated in the textbooks, see for example
\cite{teitelboim1992quantization}. This deviation has an impact on
basic statements of gauge theory. Notice the second Noether theorem,
which connects gauge symmetry of the action with Noether identities
between Lagrangian equations. We can mention two assumptions implied
by the theorem: (i) the gauge parameters are arbitrary functions of
$x$; (ii) any on-shell vanishing local quantity\footnote{By local
quantity we mean the function of space-time coordinates, fields, and
their derivatives of finite order.} reduces to a linear combination
of the l.h.s. of Lagrangian equations and their derivatives.  The
first assumption is obviously invalid once the symmetry is unfree.
The second one is also inevitably violated for the case of unfree
gauge symmetry as it is explained in the articles
\cite{KAPARULIN2019114735}, \cite{Kaparulin2019}. Let us rephrase
the violation of the second assumption: the local quantities
$\tau_a$ exist such that vanish on-shell, while they cannot be
expanded in the l.h.s. of Lagrangian equations with local
coefficients:
\begin{equation}\label{tau-a}
    \exists\, \tau_a (\phi): \quad \tau_a\approx 0\,, \quad \tau_a (\phi)\neq
    K_a^i (\phi)\partial_i S \, .
\end{equation}
Here, we use the condensed notation. The condensed indices $a,i$
include space-time point $x$ and discrete labels. Summation over
condensed indices includes integration over space-time, $\partial_i
S(\phi)$ is a variational derivative of the action $S(\phi)$  by the
field $\phi^i$, and the symbol $\approx$ means on-shell equality.
So, violation of (ii) means that ideal $I$ of on-shell vanishing
local quantities is not spanned by the l.h.s. of Lagrangian
equations $\partial_i S=0$. The local quantities $\tau\in I$
(\ref{tau-a}) are called \emph{completion functions}. The generating
set of ideal $I$ includes l.h.s. of Lagrangian equations \emph{and}
a number of completion functions. In slightly different wording, any
on-shell vanishing local quantity $T(\phi)$ is spanned off-shell by
field equations and completion functions with the local
expansion coefficients:
\begin{equation}\label{I}
    T(\phi)\approx 0 \quad\Leftrightarrow\quad
    T(\phi)=T^i(\phi)\partial_i S(\phi)+ T^a(\phi)\tau_a(\phi) \, .
\end{equation}
The identities can exist between the Lagrangian equations and
completion functions,
\begin{equation}
\label{GI} \displaystyle
\Gamma_\alpha^i(\phi)\partial_iS(\phi)+\Gamma_\alpha^a(\phi)\tau_a(\phi)\equiv0\,
,
\end{equation}
where all the coefficients $\Gamma(\phi)$ are local. These relations
can be understood as modification of the usual Noether identities
for the case when the theory admits completion functions. Upon not
quite restrictive regularity assumptions (see in
\cite{KAPARULIN2019114735}, \cite{Kaparulin2019}), the operators
 $\Gamma_\alpha^a(\phi)$, being the coefficients at completion functions, can
admit at maximum a finite dimensional kernel:
\begin{equation}\label{KerG}
\Gamma^a_\alpha (\phi)u_a=0 \quad \Rightarrow\quad u_a\in
M=\textmd{Ker}\,\Gamma^a_\alpha,\quad \dim{M}=n\in\mathbb{N} \, .
\end{equation}
The kernel $M$ is understood as a moduli space of the field theory.
Elements of $M$ are  parameterized by finite number of constant
parameters $\Lambda$. Being parameterized by constants, the elements
of $M$ can explicitly depend on the space-time point $x$. From the
viewpoint of modified Noether identities (\ref{GI}), the completion
functions $\tau_a$ are defined modulo the kernel $M$ (\ref{KerG}).
Specific element of the kernel is defined by the asymptotics of the
fields, as $\tau_a$ should vanish on-shell everywhere, including
boundary. From this perspective, the existence of completion
functions (\ref{tau-a}) can be considered as a consequence of
modified Noether identities (\ref{GI}) rather than a cause. Once the
kernel of $\Gamma^a_\alpha$ is finite, the identities (\ref{GI})
mean that the local quantities $\tau_a$ reduce on-shell to a
specific $\Lambda$-dependent function of $x$. This function can be
subtracted from $\tau$, so the completion functions vanish on-shell.
On the other hand, $\Gamma^a_\alpha$  is a differential operator,
and it does not have inverse in the class of differential operators,
as the kernel exists. Once $\Gamma^a_\alpha$ is not locally
invertible, completion function $\tau_a(\phi)$, being on-shell
vanishing local quantity, cannot be expressed from the identities
(\ref{GI}) as a linear combination of Lagrangian equations with
local coefficients. In this sense, the identities (\ref{GI}) lead to
existence of completion functions (\ref{tau-a}).

Be modified Noether identity (\ref{GI}) a consequence of existence
of completion functions (\ref{tau-a}), or vice versa, anyway, it
means that the action $S(\phi)$ enjoys unfree gauge symmetry. The
unfree gauge transformation is defined by the coefficients
$\Gamma^i_\alpha$ of the identities (\ref{GI}),
\begin{equation}
\label{gst} \displaystyle
\delta_\varepsilon\phi^i=\Gamma^i_\alpha(\phi)\varepsilon^\alpha\, ,
\end{equation}
while the operators $\Gamma^a_\alpha$  define the equations
constraining  gauge parameters:
\begin{equation}
\label{gpc} \displaystyle \Gamma_\alpha^a(\phi)\varepsilon^\alpha=0\,.
\end{equation}
Let us mention the terminology: operators $\Gamma^i_\alpha$, being
the coefficients at Lagrangian equations in modified Noether
identities (\ref{GI}), are understood as unfree gauge symmetry
generators, while $\Gamma^a_\alpha$, being the coefficients at the
completion functions in (\ref{GI}), are considered as  operators of
gauge parameter constraints. Given the identities (\ref{GI}), the
transformation (\ref{gst}) leaves the action intact off-shell once
the parameters obey conditions (\ref{gpc}):
\begin{equation}\label{USoA}
    \delta_\varepsilon S(\phi)
    \equiv \partial_i
    S(\phi)\Gamma^i_\alpha\varepsilon^\alpha\equiv-\tau_a\Gamma^a_\alpha\varepsilon^\alpha=0 \, .
\end{equation}
In this way, we see that unfree gauge symmetry is a consequence  of
modified Noether identities (\ref{GI}). Proceeding from this
observation, we can find the Hamiltinian counterpart of the unfree
gauge symmetry. It is sufficient to find the modified Noether
identities (\ref{GI}) for the Hamiltonian equations with
constraints, and the equations for gauge parameters (\ref{gpc}) are
immediately identified. This is done in the next section.

Let us briefly explain the modification of the Faddeev-Popov (FP)
ansatz needed to account for the unfree gauge symmetry. The
modification is proposed in reference \cite{KAPARULIN2019114735},
where one can find a more detailed exposition of the method. In the
section 4, we deduce this modified ansatz from the BFV-BRST
formalism.

The ghosts assigned to the unfree gauge transformations (\ref{gst})
are assumed to obey equations
\begin{equation}\label{C-constr}
\displaystyle \Gamma_\alpha^a(\varphi)C^\alpha=0\,, \qquad \text{gh}\left(C^\alpha\right)=1\,, \qquad \epsilon\left(C^\alpha\right)=1\,,
\end{equation}
where $\Gamma_\alpha^a(\varphi)$ are the operators of gauge parameter
constraints (\ref{gpc}). Let us impose independent gauges
$\chi^I(\phi)$. The index $I$ is condensed, so it includes the space
coordinates $x^\mu$. The dimension of digital part of the index
should be equal to the number of unconstrained gauge
parameters\footnote{In the next section, we explain the number of
gauge conditions from the Hamiltonian perspective.}. Once we use
independent gauge-fixing conditions,
 the number of unfree gauge parameters will exceed the number of gauges, so FP matrix will be rectangular,
\begin{equation}\label{FPM}
   \frac{\delta_\varepsilon\chi^I}{\delta\varepsilon^\alpha} =\Gamma^i_\alpha(\phi)\partial_i\chi^I (\phi)\,.
\end{equation}
Given the admissible gauge fixing conditions, the anti-ghosts
\begin{equation}\label{barC}
    \bar{C}_I\,, \qquad \text{gh}\left(\bar{C}_I\right)=-1, \qquad \epsilon\left(\bar{C}_I\right)=1
    \end{equation}
 are assigned to $\chi^I(\phi)$.
The FP ansatz for path integral is adjusted to the case of
unfree gauge symmetry in the following way:
\begin{equation}\label{ZFP}
Z= \int [d\Phi]\exp\Big\{{\frac{i}{\hbar}S_{FP}(\varphi)}\Big\} \, , \qquad \Phi=\{\phi^i, \pi_I, C^\alpha,\bar{C}_I, \bar{C}_a\} \, ,
\end{equation}
\begin{equation}
\displaystyle \text{gh}\left(\bar{C}_a\right)=-1,\qquad \epsilon\left(\bar{C}_I\right)=1\,; \qquad \text{gh}\left(\pi_I\right)=\epsilon\left(\pi_I\right)=0\,,
\end{equation}
where the FP action reads
\begin{equation}\label{FP-act}
    S_{FP}= S(\phi) + \pi_I\chi^I(\phi)+
    \bar{C}_I\Gamma^i_\alpha(\phi)\partial_i\chi^I(\phi)C^\alpha +
    \bar{C}_a\Gamma^a_\alpha (\phi) C^\alpha \, .
\end{equation}
The Fourier multipliers $\bar{C}_a$ to the ghost constraints
$\Gamma^a_\alpha(\phi)C^\alpha=0$ can be considered as anti-ghosts, on
equal footing with the anti-ghosts $\bar{C}_I$ assigned to the gauge-fixing conditions $\chi^I(\phi)$. In the section 4, we shall see
that these anti-ghosts naturally arise from the Hamiltonian BFV-BRST
formalism.

Let us exemplify the above generalities about unfree gauge symmetry
by the case of UG.  Consider the unimodular metrics $g_{\mu\nu}(x),
\, \text{det}\,g=-1$, in $d=4$. The usual explanations of gauge symmetry in
UG proceed from the idea that the symmetry is a diffeomorphism
consistent with unimodularity condition. This imposes the
transversality equation on the parameter. We go another way,
following the procedure above, and we shall see the same result.

Lagrangian equations of UG read:
\begin{equation}\label{UGE}
\frac{\delta S[g]}{\delta g_{\mu\nu}}\equiv
R^{\mu\nu}-\frac{1}{4}g^{\mu\nu}R\approx 0 \, , \quad S=\int d^4x \,
R \, .
\end{equation}
Taking divergence of the equations, and making use of Bianchi
identity, we get
\begin{equation}\label{UGediv}
\nabla_\nu \frac{\delta S[g]}{\delta g_{\mu\nu}}\equiv \nabla^\mu
R\approx 0 \, .
\end{equation}
 \noindent
 Unlike GR, the divergence of the field equations does not identically vanish. Once $\partial_\mu R\approx 0$, the scalar curvature is
 an on-shell constant, $R\approx\Lambda=const$, where specific value of $\Lambda$ is
 defined by asymptotics of $g_{\mu\nu}$. So we have the modular parameter $\Lambda$, and completion
 function $\tau\equiv R-\Lambda\approx 0$.  Obviously, $\tau$ cannot
 be represented as  linear combination of  equations (\ref{UGE})
 and their derivatives, so it is a completion function indeed.
 Then, we get modified Noether identities (\ref{GI}) for UG:
\begin{equation}\label{GIUG}
\nabla_\nu \frac{\delta S[g]}{\delta g_{\mu\nu}}-
\nabla^\mu\tau\equiv 0\,.
\end{equation}
This allows us to identify the unfree gauge symmetry transformations
(\ref{gst}), and the gauge parameter constraints (\ref{gpc}):
\begin{equation}\label{gpcUG}
    \delta_\varepsilon g_{\mu\nu}
    =\nabla_\mu\varepsilon_\nu+\nabla_\nu\varepsilon_\mu\, ,\quad
     \nabla_\mu\varepsilon^\mu=0 \, .
     \end{equation}
We can also mention one more example of completion function noticed
in literature concerning Maxwell-like higher spin field theory
\cite{Campoleoni2013}. In this theory, the double divergence of the
tracefull second-rank tensor vanishes on-shell,
$\partial_\mu\partial_\nu\varphi^{\mu\nu}\approx 0$, while it does
not reduce to the l.h.s. of the field equations and their
derivatives. This fact is emphasized in the article
\cite{Francia2017}.

In the end of this section, we provide, without proof, a receipt for
covariant degree of freedom (DoF) counting in the theories with
unfree gauge symmetry. In so doing, we assume that the Lagrangian
equations are involutive in the sense that they do not admit lower
order \emph{differential} consequences. The receipt can be deduced
along the same lines as explained in the article
\cite{Kaparulin:2012px} for the gauge theories without constraints
on gauge parameters.

DoF number is calculated as follows:
\begin{equation}\label{DoF}
\displaystyle
N_{\text{DoF}}=n_{e}o_{e}-n_{s}o_{s}-n_{i}o_{i}+n_{c}o_{c}\,,
\end{equation}
where $n_{e}$, $n_{s}$, $n_{i}$. $n_{c}$ are the numbers, and
$o_{e}$, $o_{s}$, $o_{i}$, $o_{c}$ are the orders of Lagrangian
equations $\displaystyle \partial_iS=0$, gauge
symmetry transformations $\displaystyle \delta_\varepsilon\varphi^i=
\Gamma^i_\alpha\varepsilon^\alpha$, gauge identities $\displaystyle
\Gamma^i_\alpha\partial_iS+\Gamma_\alpha^a\tau_a=0$, and constraints
$\displaystyle \Gamma_\alpha^a\varepsilon^\alpha=0$, respectively. The order
$o_{e}$ is defined by the highest order derivative in EoMs, $o_{s}$
is the order of gauge symmetry differential operator. The order of
gauge identity, $o_{i}$, is a sum of $o_{s}$ and $o_{e}$, and
$o_{c}$ is a sum of the order of constraint operator
$\Gamma_\alpha^a$ and $o_{s}$\,.

Let us exemplify the DoF number count (\ref{DoF}) by the case of UG
in $d=4$. We have nine equations of the second order (\ref{UGE}),
$n_{e}=9$, $o_{e}=2$. There are four gauge symmetry transformations
of the first order, and one first-order equation imposed on the gauge
parameters (\ref{gpcUG}),  so $n_{s}=4, o_{s}=1, n_{c}=1,
o_{c}=1+1=2$. There exist four gauge identities (\ref{GIUG}), $n_i=4$,
 of the third order ($o_{i}=1+2=3$). So, according to
(\ref{DoF}), UG has four degrees of freedom by phase-space count,
which corresponds to two ``Lagrangian" DoF.


\section{Constrained Hamiltonian formalism: \\ higher order constraints, modular parameters, and unfree gauge symmetry.}
\noindent Any action functional can be brought to equivalent
Hamiltonian form with primary constraints:
\begin{equation}\label{Sgen}
\displaystyle S=\int dt \big(p_i\dot{q}^i-H_T(q,p,\lambda)\big)\,, \qquad H_T(q,p,\lambda)=H(q,p)+\lambda^{\alpha_1} \overset{(1)}T{}_{\alpha_1}(q,p)\,,
\end{equation}
where $q^i, p_i$ are canonical variables, and $\lambda^{\alpha_1}$
are Lagrange multipliers. All these variables can be viewed as the
fields $\phi=(q,p,\lambda)$, and then we can apply the general
consideration of the previous section to the action (\ref{Sgen}). As
explained in the previous section, the unfree gauge symmetry
(\ref{gst}), (\ref{gpc}) is caused by modified Noether identities
(\ref{GI}) which involve, besides the original Lagrangian equations
$\partial_i S(\phi)=0$ and gauge generators $\Gamma^i_\alpha$ two
more ingredients: completion functions $\tau_a(\phi)$  and operators
of gauge parameter constraints $\Gamma^a_\alpha$. The key point in
finding the unfree gauge symmetry of any action functional is to
find a modified Noether identities (\ref{GI}) involving the operator
$\Gamma^a_\alpha$  with a finite kernel (\ref{KerG}). Once the
identities are found, the coefficients at the equations define the
gauge generators, while the operators $\Gamma^a_\alpha$ give the
equations imposed on the gauge parameters. Hamiltonian action
(\ref{Sgen}), due to the canonical structure, is very convenient for
algorithmically deducing modified Noether identities (\ref{GI}). The
idea is quite simple: we apply the Dirac-Bergmann algorithm of
iterating constraints. We assume that no Lagrange multiplier is
fixed, so all the constraints are first-class. In the local field
theory, the algorithm should terminate in a finite number of
iterations. Termination of the algorithm is a (modified) Noether
identity. Once the modified Noether identities (\ref{GI}) are
established, one can find the gauge transformation for the fields
$\phi=(q,p,\lambda)$ by identifying the coefficients at the
corresponding equations, while the gauge parameter constraints are
defined by the coefficient at the completion functions in the
identity. As one can guess, the roles of completion functions  are
plaid in Hamiltonian formalism by secondary constraints of all
generations. For the case when the sequence of constraints
terminates at the secondary constraints, without tertiary and higher
order ones, this program has been already implemented in the article
\cite{Abakumova:2019uoo}. Here we consider the general case. When
the secondary constraints lead to the higher order ones, and the
involution coefficients include differential operators with finite
kernel, this can lead, in general, to explicit dependence of kernel
elements on space-time coordinates $x$. Through this mechanism, the
explicit time dependence can arise in the higher order constraints
even if the original action is translation-invariant. The explicit
$x$-dependence of secondary constraints is due to the field
asymptotics which is defined by modular parameters.

Let us consider iteration of secondary constraints to deduce
Hamiltonian form of identity (\ref{GI}), and get in this way the
unfree gauge symmetry (\ref{gst}), (\ref{gpc}) for Hamiltonian
action (\ref{Sgen}). EoM's read:
\begin{equation}\label{HEs}
\begin{array}{l}
\displaystyle \frac{\delta S}{\delta p_i}\equiv\dot{q}^i-\{q^i\,,H_T(q,p,\lambda)\}=0\,,
\\[3mm]
\displaystyle \frac{\delta S}{\delta q^i}\equiv-\dot{p}_i+\{p_i\,,H_T(q,p,\lambda)\}=0\,;
\end{array}
\end{equation}
\vspace{-3mm}
\begin{equation}\label{HCs}
\displaystyle \frac{\delta S}{\delta
\lambda^{\alpha_1}}\equiv-\overset{(1)}T{}_{\alpha_1}(q,p)=0\,.
\end{equation}
Following the Dirac-Bergmann algorithm, we take time derivative of
primary constraints (\ref{HCs}) and combine it with the evolutionary
equations (\ref{HEs}) to exclude the time derivatives. The result is
at most linear in $\lambda$. As the multipliers remain indefinite,
all the coefficients at $\lambda$ should be considered as on-shell
vanishing, so the derivative of the primary constraints reduces to
the combination of primary and secondary constraints:
\begin{equation}\label{T1gen}
\displaystyle \frac{d}{dt}\overset{(1)}T_{\alpha_1}(q,p)=\{\overset{(1)}T_{\alpha_1}(q,p)\,,H_T(q,p,\lambda)\}=\overset{(1)}V{}_{\alpha_1}^{\beta_1}(q,p,\lambda)\overset{(1)}T_{\beta_1}(q,p)+\overset{(1)}\Gamma{}_{\alpha_1}^{\beta_{2}}(q,p,\lambda)\overset{(2)}T_{\beta_{2}}(q,p)\,.
\end{equation}
Unfree gauge symmetry corresponds to the case when the structure
coefficient
$\overset{(1)}\Gamma{}_{\alpha_1}^{\beta_{2}}(q,p,\lambda)$ is a
differential operator with finite kernel (\ref{KerG}). This includes
the case of zero kernel, while no inverse exists for
$\overset{(1)}\Gamma$ in the class of differential operators. This
has been first noticed in Ref. \cite{Abakumova:2019uoo}, though this
article assumed no higher order constraints appear. Relation
(\ref{T1gen}) defines secondary constraints $\overset{(2)}T$ modulo
kernel of $\overset{(1)}\Gamma$. The kernel is parameterized by
finite set of constant modular parameters $\Lambda$. The elements of
the kernel can be specific $\Lambda$-dependent functions of
space-time point $x$. The latter fact means that $\overset{(2)}T$
can be explicitly time-dependent,
\begin{equation}\label{T2}
\overset{(2)}{T}_{\beta_{2}}(q,p,\Lambda, t)={T}_{\beta_{2}}(q,p)+
{u}_{\beta_{2}}(\Lambda, t, q,p)\,, \qquad
\overset{(1)}\Gamma{}_{\alpha_1}^{\beta_{2}}{u}_{\beta_{2}}(\Lambda,
t, q,p)=0\, .
\end{equation}
Further examination of the stability of the secondary constraints
has to account for the possible explicit time-dependence which can
originate from the kernel of $\overset{(1)}\Gamma$. The kernel
depends, in its own turn, on the asymptotics of the fields.

Consider now the sequence of $n$ stability conditions of constraints
labeled by index $k, \, k=2,\ldots, n$. The time derivatives of
secondary constraints should vanish on-shell that leads to tertiary
constraints, etc. Stability of the $l$-order constraints
$\overset{(l)}T$ leads to $\overset{(l+1)}T$:
 \vspace{-3mm}
\begin{equation}\label{Tkgen}
\begin{array}{l}
\displaystyle \frac{d}{dt}\overset{(l)}T_{\alpha_l}(q,p)=\frac{\partial}{\partial t}\overset{(l)}T_{\alpha_l}(q,p)+\{\overset{(l)}T_{\alpha_l}(q,p)\,,H_T(q,p,\lambda)\}=\\[3mm]
\displaystyle=\sum\limits_{m=1}^{l}\overset{(l)}V{}_{\alpha_l}^{\beta_m}(q,p,\lambda)\overset{(m)}T_{\beta_m}(q,p)+\overset{(l)}\Gamma{}_{\alpha_l}^{\beta_{l+1}}(q,p,\lambda)\overset{(l+1)}T_{\beta_{l+1}}(q,p)\,, \quad l=2,\ldots,n-1\,.
\end{array}
\end{equation}
The coefficients $\overset{(l)}\Gamma$ at the constraints of next
generation $\overset{(l+1)}T$  are the differential operators with a
finite kernel. Therefore, constraints of $(l+1)$-st generation are
defined modulo the kernel elements much like the secondary ones
(\ref{T2}). In general, the kernel is different for different $l$'s.
The algorithm terminates when no further constraints appear:
\vspace{-3mm}
\begin{equation}\label{Tngen}
\displaystyle \frac{d}{dt}\overset{(n)}T_{\alpha_n}(q,p)=\frac{\partial}{\partial t}\overset{(n)}T_{\alpha_n}(q,p)+\{\overset{(n)}T_{\alpha_n}(q,p)\,,H_T(q,p,\lambda)\}=\sum\limits_{m=1}^{n}\overset{(n)}V{}_{\alpha_n}^{\beta_m}(q,p,\lambda)\overset{(m)}T_{\beta_m}(q,p)\,.
\end{equation}
\vspace{-5mm}\\
Note, that constraints $\overset{(k)}T_{\alpha_k}, k=2,\ldots,n,$
contain modular parameters defined by asymptotics of the field and
can be explicitly time-dependent. Once $\Gamma$'s are differential
operators, the secondary constraints of all generations
(\ref{T1gen}), (\ref{Tkgen}) are \emph{not differential}
consequences of original variational equations (\ref{HEs}),
(\ref{HCs}), while they vanish on-shell, so they are completion
functions (\ref{tau-a}).

Notice that all the structure functions $V,\Gamma$ in relations
(\ref{T1gen}), (\ref{Tkgen}), (\ref{Tngen}) are at most linear in $\lambda$, so it
is useful to introduce separate notation for the coefficients at
$\lambda$'s and $\lambda$-independent terms:
\begin{equation} \displaystyle
\overset{(r)}V{}_{\alpha_r}^{\beta_s}(q,p,\lambda)=V{}_{\alpha_r}^{\beta_s}(q,p)+U{}_{\alpha_r\gamma_1}^{\beta_s}(q,p)\lambda^{\gamma_1}\,,
\quad r,s=1,\ldots,n\,;
\end{equation}
\begin{equation}
\overset{(r)}\Gamma{}_{\alpha_r}^{\beta_{r+1}}(q,p,\lambda)=\Gamma{}_{\alpha_r}^{\beta_{r+1}}(q,p)+U{}_{\alpha_r\gamma_1}^{\beta_{r+1}}(q,p)\lambda^{\gamma_1}\,, \quad r=1,\ldots,n\,, \quad r+1\leq n\,.
\end{equation}

Once the secondary constraints $\overset{(k)}T, \, k=2,\ldots, n$ of
all generations play the role of completion functions (\ref{tau-a}),
the relations of the Dirac-Bergmann algorithm (\ref{T1gen}),
(\ref{Tkgen}), (\ref{Tngen}) can be assembled into the modified
Noether identities (\ref{GI}):
\begin{equation}
\label{HGI1} \displaystyle
\{\overset{(1)}T_{\alpha_1}\,,q^i\}\frac{\delta S}{\delta
q^i}+\{\overset{(1)}T_{\alpha_1}\,,p_i\}\frac{\delta S}{\delta
p_i}+\big(\delta_{\alpha_1}^{\beta_1}\frac{d}{dt}-\overset{(1)}V{}_{\alpha_1}^{\beta_1}(q,p,\lambda)\big)\frac{\delta
S}{\delta
\lambda^{\beta_1}}+\overset{(1)}\Gamma{}_{\alpha_1}^{\beta_2}(q,p,\lambda)\overset{(2)}T_{\beta_2}\equiv
0\,;
\end{equation}
\begin{equation}\label{HGIk}
\begin{array}{c}
\displaystyle \{\overset{(l)}T_{\alpha_l}\,,q^i\}\frac{\delta S}{\delta q^i}+\{\overset{(l)}T_{\alpha_l}\,,p_i\}\frac{\delta S}{\delta p_i}-\overset{(l)}V{}_{\alpha_l}^{\beta_1}(q,p,\lambda)\frac{\delta S}{\delta \lambda^{\beta_1}}+\sum\limits_{m=2}^{l-1}\overset{(l)}V{}_{\alpha_l}^{\beta_m}(q,p,\lambda)\overset{(m)}T_{\beta_m}\,+\\[3mm]
\displaystyle
+\,\big(-\delta_{\alpha_l}^{\beta_l}\frac{d}{dt}+\overset{(l)}V{}_{\alpha_l}^{\beta_l}(q,p,\lambda)\big)\overset{(l)}T_{\beta_l}+\overset{(l)}\Gamma{}_{\alpha_l}^{\beta_{l+1}}(q,p,\lambda)\overset{(l+1)}T_{\beta_{l+1}}\equiv
0\,, \quad l=2,\ldots,n-1\,;
\end{array}
\end{equation}
\begin{equation}\label{HGIn}
\begin{array}{c}
\displaystyle \{\overset{(n)}T_{\alpha_n}\,,q^i\}\frac{\delta S}{\delta q^i}+\{\overset{(n)}T_{\alpha_n}\,,p_i\}\frac{\delta S}{\delta p_i}-\overset{(n)}V{}_{\alpha_n}^{\beta_1}(q,p,\lambda)\frac{\delta S}{\delta \lambda^{\beta_1}}\,+\\[3mm]
\displaystyle
+\,\sum\limits_{l=2}^{n-1}\overset{(n)}V{}_{\alpha_n}^{\beta_l}(q,p,\lambda)\overset{(l)}T_{\beta_l}+\big(-\delta_{\alpha_n}^{\beta_n}\frac{d}{dt}+\overset{(n)}V{}_{\alpha_n}^{\beta_n}(q,p,\lambda)\big)\overset{(n)}T_{\beta_n}\equiv
0\,.
\end{array}
\end{equation}
The coefficients at the variational equations in the identities
(\ref{GI}) define unfree gauge variations (\ref{gst}) of
corresponding variables, while the coefficients at completion
functions define the constraints imposed on the gauge parameters
(\ref{gpc}). Given the modified Noether identities in the
Hamiltonian form (\ref{HGI1}), (\ref{HGIk}), (\ref{HGIn}), with
$q,p,\lambda$ being the fields, and the secondary constraints
$\overset{(k)}T$ being the completion functions, we arrive at the
Hamiltonian form of the unfree gauge symmetry:
\begin{equation} \label{ugtqp}
\displaystyle \delta_\varepsilon
O(q,p)=\sum\limits_{r=1}^n\{O\,,\overset{(r)}T_{\alpha_r}\}\varepsilon^{\alpha_r}\,;
\end{equation}
\vspace{-3mm}
\begin{equation} \label{ugtlambda}
\displaystyle
\delta_\varepsilon\lambda^{\alpha_1}=\big(\delta^{\alpha_1}_{\beta_1}\frac{d}{dt}+\overset{(1)}V{}_{\beta_1}^{\alpha_1}(q,p,\lambda)\big)\varepsilon^{\beta_1}+\sum\limits_{k=2}^n\overset{(k)}V{}_{\beta_k}^{\alpha_1}(q,p,\lambda)\varepsilon^{\beta_k}\,,
\end{equation}
while equations constraining gauge parameters (\ref{gpc}) read
\begin{equation}\label{ugtc1}
\begin{array}{c}
\displaystyle \big(\delta_{\beta_l}^{\alpha_l}\frac{d}{dt}+\overset{(l)}V{}_{\beta_l}^{\alpha_l}(q,p,\lambda)\big)\varepsilon^{\beta_l}+\sum_{m=l+1}^n\overset{(m)}V{}_{\beta_m}^{\alpha_l}(q,p,\lambda)\varepsilon^{\beta_m}+\overset{(l-1)}\Gamma{}_{\beta_{l-1}}^{\alpha_l}(q,p,\lambda)\varepsilon^{\beta_{l-1}}=0\,,
\end{array}
\end{equation}
where $l=2,\ldots,n-1$,
\begin{equation}\label{ugtc2}
\displaystyle \big(\delta_{\beta_n}^{\alpha_n}
\frac{d}{dt}+\overset{(n)}V{}_{\beta_n}^{\alpha_n}(q,p,\lambda)\big)\varepsilon^{\beta_n}+\overset{(n-1)}\Gamma{}_{\beta_{n-1}}^{\alpha_n}(q,p,\lambda)\varepsilon^{\beta_{n-1}}=0\,.
\end{equation}
As one can see, the gauge transformations are generated by the
constraints of all generations (\ref{ugtqp}), (\ref{ugtlambda}),
while corresponding gauge parameters are bound by the differential
equations (\ref{ugtc1}), (\ref{ugtc2}). One can verify by direct
computation that transformations (\ref{ugtqp}), (\ref{ugtlambda})
leave original Hamiltonian action (\ref{Sgen}) intact. Given
involution relations of  Hamiltonian and constraints (\ref{T1gen}),
(\ref{Tkgen}), (\ref{Tngen}), the gauge variation (\ref{ugtqp}),
(\ref{ugtlambda}) of the action reads:
\begin{equation*}
\displaystyle \delta_\varepsilon S\equiv \int dt
\Big\{\sum\limits_{l=2}^{n-1}\bigg[\big(\delta_{\beta_l}^{\alpha_l}\frac{d}{dt}+\overset{(l)}V{}_{\beta_l}^{\alpha_l}\big)\varepsilon^{\beta_l}+\sum_{m=l+1}^n\overset{(m)}V{}_{\beta_m}^{\alpha_l}\varepsilon^{\beta_m}+\overset{(l-1)}\Gamma{}_{\beta_{l-1}}^{\alpha_l}\varepsilon^{\beta_{l-1}}\bigg]\overset{(l)}T_{\alpha_l}\,+
\end{equation*}
\vspace{-3mm}
\begin{equation}\label{S-invar}
\displaystyle\qquad\qquad+\,\bigg[\big(\delta_{\beta_n}^{\alpha_n}
\frac{d}{dt}+\overset{(n)}V{}_{\beta_n}^{\alpha_n}\big)\varepsilon^{\beta_n}+\overset{(n-1)}\Gamma{}_{\beta_{n-1}}^{\alpha_n}
\varepsilon^{\beta_{n-1}}\bigg]\overset{(n)}T_{\alpha_n}-\frac{1}{2}\frac{d}{dt}\left(\sum\limits_{r=1}^n\overset{(r)}T_{\alpha_r}\varepsilon^{\alpha_r}\right)\Big\}=0\,.
\end{equation}
By virtue of equations (\ref{ugtc1}), (\ref{ugtc2}), imposed on
the gauge parameters, the integrand reduces to the total derivative.

Let us discuss the structure of equations (\ref{ugtc1}),
(\ref{ugtc2}) constraining  gauge parameters. To demonstrate key
features of the equations, consider the toy model such that has only
one constraint of each generation, so no indices $\alpha_k$ are
needed. The next simplification is that all the constraints commute.
So, the involution relations (\ref{T1gen}), (\ref{Tkgen}),
(\ref{Tngen}) get a simple form:
\begin{equation}\label{OneTcase}
    \displaystyle \{\overset{(1)}T,H\}= \overset{(1)}\Gamma\,\overset{(2)}T\, ,
    \quad \{\overset{(l)}T,H\}=
    \overset{(l)}\Gamma\,\overset{(l+1)}T\,,  \quad
    \{\overset{(n)}T,H\}=0\,; \qquad
    \{\overset{(r)}T,\overset{(s)}T\}=0 \,,
\end{equation}
where $l=2,\ldots,n-1;\,\, r,s=1,\ldots,n$.
Given the involution relations, gauge transformations (\ref{ugtqp}),
(\ref{ugtlambda}) read:
\begin{equation}\label{ugt-example}
    \displaystyle \delta_\epsilon O =\sum\limits_{r=1}^{n}\{O\,,\overset{(r)}T\}\varepsilon^{r}\,,\qquad
    \delta_\epsilon \lambda=\dot{\varepsilon}^1\,.
\end{equation}
The equations (\ref{ugtc1}), (\ref{ugtc2}) constraining gauge
parameters $\varepsilon^r$ read:
\begin{equation}\label{eps-eq-example}
\displaystyle
\dot{\varepsilon}^{r+1}+\overset{(r)}\Gamma\varepsilon^{r}=0\,,
\quad r=1,\ldots,n-1\,.
\end{equation}
If operators $\overset{(r)}\Gamma, \, r=1,\ldots, n-1$, were all
invertible in the class of differential operators, one could express
all the gauge parameters $\varepsilon^r$ as the derivatives of the
last one:
\begin{equation}\label{eps-res}
    \varepsilon^{r}=\left(\overset{(r)}\Gamma\right)^{-1}\frac{d}{dt}\left(\overset{(r+1)}\Gamma\right)^{-1}\ldots\frac{d}{dt}
    \left(\overset{(n-1)}\Gamma\right)^{-1}\frac{d}{dt}\varepsilon^n
    \,.
\end{equation}
Relation (\ref{eps-res}) is a general solution for  equations
(\ref{eps-eq-example}). Given the solution, one can substitute all the
gauge parameters $\varepsilon^r, \, r=1, \ldots, n-1$,  in terms of
the unique unconstrained parameter $\varepsilon^n$, into the gauge
transformation (\ref{ugtqp}), (\ref{ugtlambda}). In this way, we
arrive at the gauge transformation without constraints on gauge
parameters but with higher derivatives of the unconstrained
parameter.  The most general case of this type, when the higher
order gauge transformation generators can be constructed for the
evolutionary equations with constraints, is considered in the
article \cite{Lyakhovich:2008hu}. The unfree gauge symmetry arises
in the example above when at least one of  operators $\Gamma$ in
involution relations (\ref{OneTcase}) does not admit inverse in the
class of differential operators. Notice the special case of this
type, when operators are non-degenerate, i.e. $\ker\Gamma=0$, while
no $\Gamma^{-1}$ exist in the class of differential operators. As
the example, we can mention the unimodular gravity with
asymptotically flat metric. The role of $\Gamma^a_\alpha$ is plaid
by partial derivative $\partial_\mu$, whose kernel is a constant. If
the fields vanish at infinity, the kernel is zero, while no local
inverse exists for the operator. In this case, the higher order
unconstrained symmetry can exist, though it is reducible. For the
linear field theories, this class of gauge parameter constraints is
described in reference \cite{FRANCIA2014248} in Lagrangian
formalism. The reducible unconstrained symmetry for this class of
nonlinear theories will be considered elsewhere.

Let us mention that the number of equations (\ref{ugtc1}),
(\ref{ugtc2}) imposed on the gauge parameters equals to the number
of secondary constraints of all generations, while the number of
gauge parameters is the number of constraints of all generations,
including primary ones. All  equations (\ref{ugtc1}), (\ref{ugtc2})
are independent, there are no identities among them, because every
equation is resolved w.r.t. the derivative of a unique gauge
parameter. Therefore, the number of independent gauge parameters
equals to the number of primary constraints. If it was possible to
locally express all the parameters in terms of independent ones and
their derivatives, like in the example above,  there would be $m_1$
independent gauge transformations, where $m_1$ is the number of
primary constraints. On the other hand, corresponding number of time
derivatives of $m_1$  independent gauge parameters $\varepsilon^n$
essentially contribute to the gauge transformations of dependent
gauge parameters $\varepsilon^r, \, r=1,\ldots, n-1$. Therefore,
overall $m$ independent parameters  and their time derivatives would
be involved in the gauge transformation (\ref{ugtqp}), where $m$ is
the total number of constraints. Hence, the on-shell gauge
invariants should Poisson-commute on-shell with the constraints of
all generations, even if the gauge symmetry is unfree. This would be
true even if $m_1$ independent higher order gauge transformations
cannot be explicitly extracted from $m$ unfree first-order
transformations in the local way. Here, we do not provide a more
rigorous justification of this observation, limiting ourselves to
the explanations  given above.

Once the unfree gauge symmetry corresponds to the higher order
symmetry with $m_1$ independent parameters, it would be sufficient
to impose $m_1$ independent gauge-fixing conditions. This number of
required gauge conditions remain the same, even if the
independent gauge parameters cannot be explicitly found from
equations (\ref{ugtc1}), (\ref{ugtc2}) in the local form. If the
gauges are imposed only on the phase-space variables, not Lagrange
multipliers, then non-degeneracy condition of the gauges
$\chi^{\alpha_1}$ reads:
\begin{equation}\label{rankchi}
    \texttt{rank}\{\chi^{\alpha_1},T_\beta\}=m_1\, ,
\end{equation}
where $T_\beta$ stands for the complete set of all constraints,
including primary, secondary, tertiary, etc., $\beta=(\beta_1,
\ldots , \beta_n)$. Once the number is different of the constraints
and gauge-fixing conditions, the non-minimal ghost sector have to be
modified in the BFV-BRST formalism for the case of unfree gauge
symmetry. This issue is considered in the next section.

\section{Hamiltonian BFV-BRST formalism for unfree gauge symmetry}
Construction of the formalism begins with introducing the minimal
sector of ghosts. Once the on-shell gauge invariants for the unfree
gauge symmetry are defined by the requirement to Poisson-commute
on-shell with the constraints of all generations, the minimal sector
is introduced along the same lines as for any first-class
constrained system \cite{teitelboim1992quantization}. Every first-class
 constraint is assigned with canonical pair of ghosts with
usual Grassmann parity and ghost number grading:
\begin{equation}
\begin{array}{c}
\displaystyle \overset{(r)}T_{\alpha_r}\,\, \rightarrow \,\,\{C{}^{\alpha_r}\,,{\bar{P}}_{\beta_r}\}=\delta_{\beta_r}^{\alpha_r}\,, \qquad \text{gh}\left(C{}^{\alpha_r}\right)=-\text{gh}\left({\bar{P}}_{\alpha_r}\right)=1\,,\\[3mm]
\displaystyle \epsilon\left(C{}^{\alpha_r}\right)=\epsilon\left(\bar{P}_{\alpha_r}\right)=1\,, \quad r=1,\ldots,n\,.
\end{array}
\end{equation}
The Hamiltonian BFV-BRST generator of minimal sector begins with the
constraints,
\begin{equation}
\displaystyle
Q_{\text{min}}=\sum_{r=1}^nC{}^{\alpha_r}T_{\alpha_r}+\ldots\,,
\qquad \text{gh}\left(Q_{\text{min}}\right)=1\,, \qquad \epsilon \left(
Q_{\text{min}}\right)=1\,,
\end{equation}
where $\ldots$ stands for $\bar{P}$-depending terms. These terms
are iteratively defined by the equation
\begin{equation}\label{QQmim}
\{Q_{\text{min}}\,,Q_{\text{min}}\}=0\,.
\end{equation}
The ghost extension of the Hamiltonian begins with the original
Hamiltonian $H$,
\begin{equation}\label{Hscript}
\displaystyle \mathcal{H}=H+\ldots\,, \qquad
\text{gh}\left(\mathcal{H}\right)=0\,, \qquad\epsilon\left(\mathcal{H}\right)=0\,.
\end{equation}
The specifics of the unfree gauge symmetry is that the completion
functions (\ref{tau-a}), and hence the secondary constraints may
depend on the space-time coordinates, even if the original
Lagrangian is $x$-independent. The $x$-dependence of the constraints
is connected with the asymptotics of the fields. Once the
constraints involve time, the BRST generator $Q_\texttt{min}$ can be
explicitly time-dependent. The explicit time dependence of
$Q_\texttt{min}$ results in appropriate modification
\cite{Batalin:1990tn} of the equation for $\mathcal{H}$\,:
\begin{equation}\label{Hscript-eq}
\displaystyle \frac{\partial}{\partial
t}Q_{\text{min}}+\{Q_{\text{min}}\,,\mathcal{H}\}=0\, .
\end{equation}
This equation defines the $\bar{P}$-dependent terms in
$\mathcal{H}$. Equation (\ref{Hscript-eq}) means that Hamiltonian
$\mathcal{H}$ is not BRST-invariant. This is a natural consequence
of the relations (\ref{Tkgen}) which mean that the original
Hamiltonian is not invariant under the unfree gauge symmetry
transformations (\ref{ugtqp}), (\ref{ugtc1}), (\ref{ugtc2}). The
Hamiltonian action (\ref{Sgen}), however, is gauge invariant, see
(\ref{S-invar}). For a similar reason, the corresponding path
integral is gauge-invariant in the BRST-BFV formalism, even though
the Hamiltonian $\mathcal{H}$, being a solution of equation
(\ref{Hscript-eq}), is not a BRST invariant. This fact is proven for
general non-stationary constrained system in the reference
\cite{Batalin:1990tn}.

Consider now the non-minimal sector for the unfree gauge theory.
Once the number of gauge fixing conditions coincides with the number
of primary constraints, the same number of non-minimal sector ghosts
is introduced,
\begin{equation}
\displaystyle
\{P{}^{\alpha_1}\,,{\bar{C}}_{\beta_1}\}=\delta^{\alpha_1}_{\beta_1}\,,
\quad
\text{gh}\left(P{}^{\alpha_1}\right)=-\text{gh}\left({\bar{C}}_{\alpha_1}\right)=1\,,
\quad
\varepsilon\left(P{}^{\alpha_1}\right)=\varepsilon\left({\bar{C}}_{\alpha_1}\right)=1\,.
\end{equation}
The Lagrange multiplier canonical pairs are introduced for primary
constraints $\overset{(1)}{T}_{\alpha_1}$ and gauge fixing conditions
$\chi^{\alpha_1}$:
\begin{equation}\label{pilambda}
\displaystyle
\{\lambda{}^{\alpha_1}\,,\pi_{\beta_1}\}=\delta^{\alpha_1}_{\beta_1}\,,
\quad \text{gh}\left(\lambda{}^{\alpha_1}\right)=\text{gh}\left(\pi_{\alpha_1}\right)=0\,,
\quad
\varepsilon\left(\lambda{}^{\alpha_1}\right)=\varepsilon\left(\pi_{\alpha_1}\right)=1\,.
\end{equation}
Complete BRST generator extends the minimal sector one in the usual
way,
\begin{equation}\label{Q-ext}
\displaystyle
Q=Q_{\text{min}}+\pi_{\alpha_1}P{}^{\alpha_1}\,.
\end{equation}
Gauge-fixing conditions involve the time derivative of Lagrange
multiplier and the function of original phase-space variables,
\begin{equation}
\label{lambda-dot-chi} \displaystyle
{\dot{\lambda}}{}^{\alpha_1}-{\chi}{}^{\alpha_1}(q,p)=0
\, .
\end{equation}
Given the gauge conditions, the gauge Fermion is introduced,
\begin{equation}\label{Psi}
    \Psi={\bar{C}}_{\alpha_1}{\chi}{}^{\alpha_1}+\lambda{}^{\alpha_1}{\bar{P}}_{\alpha_1}\,,
\end{equation}
and gauge-fixed Hamiltonian is defined by the usual rule,
\begin{equation}\label{H-psi}
\displaystyle H_\Psi=\mathcal{H}+\{Q\,,\Psi\}\,.
\end{equation}
This Hamiltonian provides conservation of the BRST generator $Q$
much like $\mathcal{H}$. The gauge-fixed BFV-BRST action reads:
\begin{equation}\label{spsi-gen}
\displaystyle S_{\text{BRST}}^\Psi=\int dt\Bigl(p_i\dot{q}^i+\pi_{\alpha_1}{\dot{\lambda}}{}^{\alpha_1}+\sum\limits_{r=1}^n{\bar{P}}_{\alpha_r}\dot{{C}}{}^{\alpha_r}+{\bar{C}}_{\alpha_1}\dot{{P}}{}^{\alpha_1}-H_\Psi\Bigr)\,.
\end{equation}
This action accounts for unfree gauge symmetry in two ways. First,
the non-minimal sector is asymmetric with the minimal one unlike the
usual BFV formalism. Second, the secondary constraints, being a part
of the BRST generator $Q$, may be explicitly time-dependent, even
though the original action does not involve time explicitly. Both of
these features do not obstruct the usual reasoning that justifies
$\Psi$-independence of the transition amplitude for this action,
\begin{equation}\label{ZPsigen}
\displaystyle Z_\Psi=\int [D\Phi]\exp\Big\{\frac{i}{\hbar}S_{\text{BRST}}^\Psi\Big\}\,,
\end{equation}
where $\Phi=\big\{q^i,p_i,\lambda{}^{\alpha_1},\pi_{\alpha_1},
C{}^{\alpha_1},{\bar{P}}_{\alpha_1},C{}^{\alpha_2},
{\bar{P}}_{\alpha_2},\ldots,C{}^{\alpha_n},{\bar{P}}_{\alpha_n},
P{}^{\alpha_1},{\bar{C}}_{\alpha_1}\big\}$.

Let us consider a theory (\ref{Sgen}) with constraints
(\ref{T1gen}), (\ref{Tkgen}), (\ref{Tngen}), with the involution
relations
\begin{equation}\label{Inv-FP}
\displaystyle
\{\overset{(r)}T_{\alpha_r}(q,p)\,,\overset{(s)}T_{\alpha_s}(q,p)\}=U_{\alpha_r\alpha_s}^{\gamma_t}(q,p)\overset{(t)}T_{\gamma_t}(q,p)\,,
\qquad r,s,t=1,\ldots,n\,.
\end{equation}
Assume that  BRST generator $Q_{\texttt{min}}$ and Hamiltonian
$\mathcal{H}$ are at most linear in the ghost momenta:
\begin{equation}\label{QFP}
\displaystyle
Q_{\text{min}}=\sum_{r=1}^nC{}^{\alpha_r}\overset{(r)}T_{\alpha_r}+\frac{1}{2}\sum\limits_{r,s,t=1}^n{C}{}^{\beta_s}{C}{}^{\alpha_r}U_{\alpha_r\beta_s}^{\gamma_t}{\bar{P}}{}_{\gamma_t}\,;
\end{equation}
\begin{equation}
\displaystyle
\mathcal{H}=H(q,p)+\sum\limits_{r=1}^{n}{C}{}^{\alpha_r}\Big(\sum\limits_{s=1}^{n}\overset{(r)}V{}^{\beta_s}_{\alpha_r}(q,p,\lambda){\bar{P}}_{\beta_s}+\overset{(r)}{\Gamma}{}^{\beta_{r-1}}_{\alpha_r}(q,p,\lambda){\bar{P}}_{\beta_{r-1}}\Big)\,.
\end{equation}
We also assume the following form of gauge-fixed Hamiltonian:
\begin{equation}\label{HFP}
\displaystyle
H_\Psi=\mathcal{H}+\{Q\,,\Psi\}=\mathcal{H}+\lambda^{\alpha_1}\overset{(1)}{T}_{\alpha_1}+\pi_{\alpha_1}{\chi}{}^{\alpha_1}+{\bar{P}}_{\alpha_1}{P}{}^{\alpha_1}+\sum\limits_{r=1}^{n}{\bar{C}}_{\alpha_1}\{{\chi}{}^{\alpha_1}\,,\overset{(r)}{T}_{\alpha_r}\}{C}{}^{\alpha_r}
\,.
\end{equation}
This is automatically true if the gauge conditions $\chi$ Poisson-commute to structure functions $U$ in the involution relations
(\ref{Inv-FP}). Given the action, path integral (\ref{ZPsigen})
reads
\begin{equation*}
\displaystyle Z_\Psi=\int[D\Phi]\exp\Big\{\frac{i}{\hbar}\int dt\Big[p_i\dot{q}^i-H(q,p)-\lambda^{\alpha_1}\overset{(1)}{T}_{\alpha_1}+\pi_{\alpha_1}(\dot{\lambda}^{\alpha_1}-{\chi}{}^{\alpha_1})\,+
\end{equation*}
\begin{equation*}
\displaystyle-\,\sum\limits_{r=1}^{n}{\bar{C}}_{\alpha_1}\{{\chi}{}^{\alpha_1}\,,\overset{(r)}{T}_{\alpha_r}\}{C}{}^{\alpha_r}+{\bar{P}}_{\alpha_1}\big(\dot{{C}}{}^{\alpha_1}+\sum\limits_{s=1}^{n}\overset{(s)}{V}{}^{\alpha_1}_{\beta_s}(q,p,\lambda)C^{\beta_s}\big)+{P}{}^{\alpha_1}\big({\bar{P}}_{\alpha_1}+\dot{{\bar{C}}}_{\alpha_1}\big)\,+
\end{equation*}
\begin{equation}\label{ZPsigen1}
\displaystyle \,+\sum\limits_{k=2}^{n}{\bar{P}}_{\alpha_k}\Big(\big(\delta_{\beta_k}^{\alpha_k}\frac{d}{dt}+\overset{(k)}V{}_{\beta_k}^{\alpha_k}\big)C^{\beta_k}
+\sum_{m=k+1}^n\overset{(m)}V{}_{\beta_m}^{\alpha_k}C^{\beta_m}+
\overset{(k-1)}\Gamma{}_{\beta_{k-1}}^{\alpha_l}C^{\beta_{k-1}}\Big)\Big]\Big\}\,,
\end{equation}
where
$\Phi=\big\{q^i,p_i,\lambda{}^{\alpha_1},\pi_{\alpha_1},C{}^{\alpha_1},{\bar{P}}_{\alpha_1},C{}^{\alpha_2},{\bar{P}}_{\alpha_2},\ldots,C{}^{\alpha_n},{\bar{P}}_{\alpha_n},P{}^{\alpha_1},{\bar{C}}_{\alpha_1}\big\}$.
Integrating in path integral (\ref{ZPsigen1}) over
$P{}^{\alpha_1}, {\bar{P}}_{\alpha_1}$, we arrive at the following
answer for the transition amplitude
\begin{equation*}
\displaystyle Z_\Psi=\int[D\Phi']\exp\Big\{\frac{i}{\hbar}\int dt\Big[p_i\dot{q}^i-H(q,p)-\lambda^{\alpha_1}\overset{(1)}{T}_{\alpha_1}+\pi_{\alpha_1}(\dot{\lambda}^{\alpha_1}-{\chi}{}^{\alpha_1})\,+
\end{equation*}
\begin{equation*}
\displaystyle -\,\sum\limits_{r=1}^{n}{\bar{C}}_{\alpha_1}\{{\chi}{}^{\alpha_1}\,,\overset{(r)}{T}_{\alpha_r}\}{C}{}^{\alpha_r}-{\bar{C}}_{\alpha_1}\big(\dot{{C}}{}^{\alpha_1}+\sum\limits_{s=1}^{n}\overset{(s)}{V}{}^{\alpha_1}_{\beta_s}(q,p,\lambda)C^{\beta_s}\big)\,+
\end{equation*}
\begin{equation}\label{Z-H-FP}
\displaystyle
\,+\sum\limits_{k=2}^{n}{\bar{P}}_{\alpha_k}\Big(\big(\delta_{\beta_k}^{\alpha_k}\frac{d}{dt}+\overset{(k)}V{}_{\beta_k}^{\alpha_k}\big)C^{\beta_k}
+\sum_{m=k+1}^n\overset{(m)}V{}_{\beta_m}^{\alpha_k}C^{\beta_m}+
\overset{(k-1)}\Gamma{}_{\beta_{k-1}}^{\alpha_l}C^{\beta_{k-1}}\Big)\Big]\Big\}\,,
\end{equation}
where
$\Phi'=\big\{q^i,p_i,\lambda{}^{\alpha_1},\pi_{\alpha_1},C{}^{\alpha_1},C{}^{\alpha_2},{\bar{P}}_{\alpha_2},\ldots,C{}^{\alpha_n},{\bar{P}}_{\alpha_n},{\bar{C}}_{\alpha_1}\big\}$.

Let us discuss the path integral (\ref{Z-H-FP}). The first line in
(\ref{Z-H-FP}) is the original action (\ref{Sgen}) and the gauge-fixing term. The second line is the FP term for the gauge
transformations (\ref{ugtqp}), (\ref{ugtlambda}). The third line
has a natural interpretation from the viewpoint of the modified FP
ansatz in Lagrangian formalism (\ref{ZFP}), (\ref{FP-act}). The
ghost momenta $\bar{P}_{\alpha_k}, \, k=2, \ldots , n$, can be viewed
as Fourier multipliers at the constraints imposed on ghosts
\begin{equation}\label{hgc}
\displaystyle \big(\delta_{\beta_k}^{\alpha_k}\frac{d}{dt}+
\overset{(k)}V{}_{\beta_k}^{\alpha_k}(q,p,\lambda)\big)C^{\beta_k}
+\sum_{m=k+1}^n\overset{(m)}V{}_{\beta_m}^{\alpha_k}(q,p,\lambda)C^{\beta_m}+
\overset{(k-1)}\Gamma{}_{\beta_{k-1}}^{\alpha_l}(q,p,\lambda)C^{\beta_{k-1}}=0\,.
\end{equation}
These ghost constraints mirror the equations imposed on gauge
parameters in Hamiltonian formalism (\ref{ugtc1}), (\ref{ugtc2}).
So, equations (\ref{hgc}) represent Hamiltonian form of the
constraints (\ref{C-constr}) imposed on the ghosts in  the case of
unfree gauge symmetry. With this regard, the path integral
(\ref{Z-H-FP}) represents the modified FP recipe (\ref{ZFP}),
(\ref{FP-act}) for the Hamiltonian action (\ref{Sgen}), gauge
symmetry (\ref{ugtqp}), (\ref{ugtlambda}), and the constraints
(\ref{ugtc1}), (\ref{ugtc2}) on the gauge parameters. So, proceeding
from the amplitude (\ref{ZPsigen}) in the general Hamiltonian
BFV-BRST formalism for unfree gauge symmetry, in the case without
higher order ghost contributions (\ref{QFP}), (\ref{HFP}), we arrive
at the modified FP path integral (\ref{ZFP}), (\ref{FP-act}).

\section{Example: traceless massless spin $s$ gauge fields}

\subsection{Lagrangian, completion functions, and unfree gauge symmetry.}
\noindent Let us consider a theory of traceless symmetric tensor
field $\varphi_{\mu_1\ldots\mu_s}\,,
\varphi^\nu{}_{\nu\mu_3\ldots\mu_s}=0$, in $d$-dimensional Minkowski
space. The metric is chosen mostly negative,
$\eta_{\mu\nu}=\text{diag}(1,-1,\ldots,-1)$. The Lagrangian reads
\cite{SKVORTSOV2008301}:
\begin{equation}\label{Lspins}
\begin{array}{c}
\displaystyle
\mathcal{L}=(-1)^s\Big(\frac{1}{2}
\partial_\nu\varphi_{\mu_1\ldots\mu_s}\partial^\nu\varphi^{\mu_1\ldots\mu_s}-\frac{s}{2}
\partial^\nu\varphi_{\nu\mu_2\ldots\mu_s}
\partial_\rho\varphi^{\rho\mu_2\ldots\mu_s}\Big)\,+\\[3mm]
\displaystyle+\,(-1)^s\frac{s}{2}\partial^\nu\big(
\varphi_{\nu\mu_2\ldots\mu_s}\partial_\rho\varphi^{\rho\mu_2\ldots\mu_s}\big)\,.
\end{array}
\end{equation}
The last term is a total divergence, so it does not contribute to
the EoMs. We include it for convenience when constructing the
Hamiltonian formalism.

The above Lagrangian describes irreducible massless spin-$s$
representation of Poincar\'e group. One of the advantages of this
form of the irreducible higher spin theory, comparing to the
Frondsdal Lagrangian \cite{Fronsdal:1978rb}, is that it does not
involve auxiliary fields. This Lagrangian can be viewed as higher
spin extension of linearized UG
\cite{Alvarez:2006uu}, \cite{Blas:2007pp}. In this section, we
utilize this model for exemplifying all the generalities about
unfree gauge symmetry considered above in this article.

The field equations for the Lagrangian (\ref{Lspins}) read:
\begin{equation}\label{EoMspins}
\displaystyle \frac{\delta
S}{\delta\varphi^{\mu_1\ldots\mu_s}}\equiv-(-1)^s\big[\square\varphi_{\mu_1\ldots\mu_s}
-s\partial_{(\mu_1}\partial^\nu\varphi_{\nu\mu_2\ldots\mu_s)}
+\frac{s(s-1)}{d+2s-4}\eta_{(\mu_1\mu_2}\partial^\nu\partial^\rho
\varphi_{\nu\rho\mu_3\ldots\mu_s)}\big]=0\, ,
\end{equation}
where round brackets $(\mu_1\ldots\mu_s)$ mean symmetrization of all
the included indices. Taking the divergence of the l.h.s., we get
the differential consequence, Cf. (\ref{UGediv}):
\begin{equation}\label{divEoMspins}
\displaystyle \partial^{\mu_1}\frac{\delta S}{\delta
\varphi^{\mu_1\ldots\mu_s}}\equiv(-1)^{s-1}\frac{d+2s-6}{d+2s-4}\big(\partial_{(\mu_1}\tau_{\mu_2\ldots\mu_{s-1})}
-\frac{2}{d+2s-6}\eta_{(\mu_1\mu_2}\partial^\lambda\tau_{\lambda\ldots\mu_{s-2})}\big)\approx0\,,
\end{equation}
where $\tau_{\mu_2\ldots\mu_{s-1}}$ is a double divergence of the
field,
\begin{equation}\label{tauspins}
\displaystyle
\tau_{\mu_1\ldots\mu_{s-2}}=\partial^\rho\partial^\nu\varphi_{\nu\rho\ldots\mu_{s-2}}\,.
\end{equation}
Relation (\ref{divEoMspins}) means that $\tau$ reduces on-shell to
the element of the kernel of first-order differential operator.
For $s=2$, $\tau$ is a scalar, and relation (\ref{divEoMspins})
means just $\partial_\mu\,\tau=0$, so $\tau$ is just on-shell
constant. In this case, the kernel is one-dimensional. For $s \geq
3$, relation (\ref{divEoMspins}) means
\begin{equation}\label{tauspinsLambda}
\displaystyle
\tau_{\mu_1\ldots\mu_{s-2}}\approx\Lambda_{\mu_1\ldots\mu_{s-2}}\,,
\end{equation}
with $\Lambda_{\mu_1\ldots,\mu_{s-2}}$ being a solution of conformal
Killing tensor equations,
\begin{equation}\label{Killing-tensor}
\displaystyle
\partial_{(\mu_1}\Lambda_{\mu_2\ldots\mu_{s-1})}-\frac{2}{d+2s-6}\eta_{(\mu_1\mu_2}\partial^\nu\Lambda_{\nu\ldots\mu_{s-2})}=0\,.
\end{equation}
The space of conformal Killing tensors is finite dimensional, so
$\tau$ is a completion function. Specific $\Lambda$ is defined by
the asymptotic behavior of the fields. For example, if $\varphi$
vanish at infinity, then $\Lambda=0$. In this most simple case,
$\tau$ still remains a non-trivial completion function as it is a
function of field derivatives off-shell, not a fixed function of
$x$. This linear function of $\partial^2\varphi$ vanishes on-shell,
while it is not a linear combination of the Lagrangian equations
(\ref{EoMspins}). We detail the case of non-vanishing $\Lambda$
below for $s=3$.

Once $\tau$ (\ref{tauspins}) is a completion function, relation
(\ref{divEoMspins}) should be understood as modified Noether
identity (\ref{GI}) because it binds  Lagrangian equations with
completion functions:
\begin{equation}\label{GIMspins}
\displaystyle \partial^{\mu_1}\frac{\delta S}{\delta
\varphi^{\mu_1\ldots\mu_s}}+(-1)^{s}\frac{d+2s-6}{d+2s-4}\big(\partial_{(\mu_1}\tau_{\mu_2\ldots\mu_{s-1})}
-\frac{2}{d+2s-6}\eta_{(\mu_1\mu_2}\partial^\lambda\tau_{\lambda\ldots\mu_{s-2})}\big)\equiv
0\, .
\end{equation}
Given the identities (\ref{GI}), it defines unfree gauge symmetry of
the action: the coefficients at EoMs define the gauge generators
(\ref{gst}), while the ones at completion functions define the
equations (\ref{gpc}) constraining the gauge parameters. In this
way, the identities (\ref{GIMspins}) define unfree gauge symmetry
\begin{equation}\label{spinsgt}
\displaystyle
\delta_\varepsilon\varphi_{\mu_1\ldots\mu_s}=s\partial_{(\mu_1}\varepsilon_{\mu_2\ldots\mu_s)}\,,
\end{equation}
where $\varepsilon_{\mu_1\ldots\mu_{s-1}}$ are traceless symmetric
gauge parameters, $\varepsilon^\nu{}_{\nu\mu_2\ldots\mu_{s-1}}=0$,
subject to the transversality conditions
\begin{equation}\label{spinsgpc}
\displaystyle \partial^\nu\varepsilon_{\nu\mu_2\ldots\mu_{s-1}}=0\,.
\end{equation}
Transformations (\ref{spinsgt}) and constraints (\ref{spinsgpc}) are
noticed in the article \cite{SKVORTSOV2008301} where the Lagrangian
(\ref{Lspins}) is proposed. The completion functions
(\ref{tauspins}), (\ref{tauspinsLambda}) are noticed here for the
first time.

\subsection{Covariant degree of freedom count.}
Let us now apply formula (\ref{DoF}) to verify DoF number of the
spin-$s$ theory (\ref{Lspins}) in explicitly covariant way. Given
the EoMs (\ref{EoMspins}), symmetry transformations (\ref{spinsgt}), gauge identities (\ref{GIMspins}),
and constraints on gauge parameters
(\ref{spinsgpc}), we can compute all the ingredients needed to count
the DoF number by the recipe (\ref{DoF}).The number of the second-order ($o_e=2$) Lagrangian equations (\ref{EoMspins}) corresponds to the number of independent components of traceless $s$-rank tensor,
\begin{equation}
\displaystyle n_{e}=
\begin{pmatrix}
d+s-1\\ s
\end{pmatrix}-
\begin{pmatrix}
d+s-3\\ s-2
\end{pmatrix}
=\frac{(d+s-3)!}{s!(d-1)!}\big(d^2+d(2s-3)-2(s-1)\big)\,.
\end{equation}\\
The number of first-order ($o_s=1$) symmetry transformations (\ref{spinsgt}) and third-order  ($o_i=1+2=3$) gauge identities (\ref{GIMspins}) equals to the number of independent components of traceless $(s-1)$-tensor,
\begin{equation}
\displaystyle n_{s}=n_{i}=
\begin{pmatrix}
d+s-1\\ s-1
\end{pmatrix}-
\begin{pmatrix}
d+s-4\\ s-3
\end{pmatrix}
=\frac{(d+s-4)!}{(s-1)!(d-1)!}\big(d^2+d(2s-5)-2(s-2)\big)\,.
\end{equation}
There exist second-order ($o_c=1+1=2$) constraints on gauge parameters (\ref{spinsgpc}), whose number coincides with the number of independent components of traceless $(s-2)$-tensor,
\begin{equation}
\displaystyle n_{c}=
\begin{pmatrix}
d+s-3\\ s-2
\end{pmatrix}-
\begin{pmatrix}
d+s-5\\ s-4
\end{pmatrix}
=\frac{(d+s-5)!}{(s-2)!(d-1)!}\big(d^2+d(2s-7)-2(s-3)\big)\,.
\end{equation}
So, the expression (\ref{DoF}) for DoF counting in case of a theory (\ref{Lspins}) reads:
\begin{equation}
\begin{array}{c}
\displaystyle N_{\text{DoF}}=\Big[\begin{pmatrix} d+s-1\\ s
\end{pmatrix}-
\begin{pmatrix}
d+s-3\\ s-2
\end{pmatrix}\Big]\cdot2
-\Big[\begin{pmatrix} d+s-1\\ s-1
\end{pmatrix}-
\begin{pmatrix}
d+s-4\\ s-3
\end{pmatrix}\Big]\cdot1\,-\\[2mm]
\displaystyle-\Big[\begin{pmatrix} d+s-1\\ s-1
\end{pmatrix}-
\begin{pmatrix}
d+s-4\\ s-3
\end{pmatrix}\Big]\cdot3
+\Big[\begin{pmatrix} d+s-3\\ s-2
\end{pmatrix}-
\begin{pmatrix}
d+s-5\\ s-4
\end{pmatrix}\Big]\cdot2\,.
\end{array}
\end{equation}
For $d=4$ this means
\begin{equation}
\displaystyle
N_{\text{DoF}}\Big|_{d=4}=(s+1)^2\cdot2-s^2\cdot1-s^2\cdot3+(s-1)^2\cdot2=4\,.
\end{equation}
Four DoF by the phase-space count corresponds to two ``Lagrangian"
modes, which is correct number for massless spin-$s$ field in $d=4$.
\noindent
\subsection{Completion functions, asymptotics, moduli space for $s=3$.}
\noindent Let us elaborate on the contribution of field asymptotics
to the completion functions in the simplest higher spin case. For
$s=3$, Lagrangian (\ref{Lspins}) and field equations (\ref{EoMspins}) read:
\begin{equation}\label{Lagrspin3}
\displaystyle
\mathcal{L}=-\Big(\frac{1}{2}
\partial_\lambda\varphi_{\mu\nu\rho}
\partial^\lambda\varphi^{\mu\nu\rho}-\frac{3}{2}
\partial^\mu\varphi_{\mu\nu\rho}
\partial_\lambda\varphi^{\lambda\nu\rho}\Big)
-\frac{3}{2}\partial^\mu\big(\varphi_{\mu\nu\rho}
\partial_\lambda\varphi^{\lambda\nu\rho}\big)\,,
\quad \varphi^\nu{}_{\nu\mu}=0\,;
\end{equation}
\begin{equation}\label{EoMspin3}
\displaystyle \frac{\delta S}{\delta \varphi^{\mu\nu\rho}}\equiv
\square\varphi_{\mu\nu\rho}-3\partial_{(\mu}\partial^\lambda
\varphi_{\lambda\nu\rho)}+\frac{6}{d+2}
\eta_{(\mu\nu}\partial^\lambda\partial^\sigma
\varphi_{\sigma\lambda\rho)}=0\,.
\end{equation}
Taking the divergence of the field equations, we get the
differential consequence
\begin{equation}\label{Conseqspin3}
\displaystyle \displaystyle \partial^{\lambda}\frac{\delta S}{\delta
\varphi^{\lambda\mu\nu}} \equiv\frac{2d}{d+2}
\Big(\partial_{(\mu}\partial^\rho\partial^\lambda\varphi_{\lambda\rho\nu)}-
\frac{1}{d}\eta_{\mu_2\mu_3}
\partial^\lambda\partial^\rho\partial^\nu\varphi_{\nu\rho\lambda}\Big)\approx
0\, .
\end{equation}
Introduce the notation
\begin{equation}\label{tauspin3}
\displaystyle
\tau_\mu=\partial^\nu\partial^\lambda\varphi_{\mu\nu\lambda}\,.
\end{equation}
Relation (\ref{Conseqspin3}) means that $\tau_\mu$  must obey on-shell
the equation for conformal Killing vector field,
\begin{equation}\label{tau-K-s3}
    \partial_\mu\tau_\nu + \partial_\nu\tau_\mu
    -\frac{2}{d}\eta_{\mu\nu}\partial^\rho\tau_\rho\approx 0\,.
\end{equation}
The general solution of the conformal Killing equation reads
\begin{equation}
\label{Lambdas3}
\partial_\mu\Lambda_\nu +\partial_\nu\Lambda_\mu -\frac{2}{d}\eta_{\mu\nu}\partial_\rho\Lambda^\rho=0
\quad\Leftrightarrow\quad
\Lambda_\mu=a_\mu+2\eta_{\mu\nu}\omega^{\nu\rho}x_\rho+ \lambda
x_\mu+b^\nu\big(2x_\mu x_\nu -\eta_{\mu\nu}x_\rho x^\rho\big)\, ,
\end{equation}
where $a_\mu, \lambda, b^\mu, \omega^{\mu\nu}=-\omega^{\nu\mu}$ are
arbitrary (integration) constants, so there are
$\displaystyle\frac{(d+2)(d+1)}{2}$ constant parameters. Relation
(\ref{tau-K-s3}) means that  $\tau_\mu$ reduces on-shell to Killing
vector (\ref{Lambdas3}):
\begin{equation}\label{ddp-lambda}
\partial_\nu\partial_\lambda\varphi^{\mu\nu\lambda}\approx\Lambda^\mu(x;\lambda,a,b,\omega)\,.
\end{equation}
Let us shift the notation (\ref{tauspin3}): $\tau_\mu=
\partial^\nu\partial^\lambda\varphi_{\mu\nu\lambda}-\Lambda_\mu(x;\lambda,a,b,\omega)
$. Then, $\tau_\mu$ vanishes on-shell,
\begin{equation}\label{taus3}
    \tau_\mu\equiv \partial^\nu\partial^\lambda\varphi_{\mu\nu\lambda}-\Lambda_\mu(x;\lambda,a,b,\omega)\approx
    0 .
\end{equation}
So, we have a function of the field derivatives such that vanishes
on-shell, while it is not a linear combination of the l.h.s. of
Lagrangian equations (\ref{EoMspin3}) and their derivatives. This
means that $\tau_\mu$ is a completion function, according to
definition (\ref{tau-a}). Relation (\ref{taus3}) can be considered
as spin-3 analogue of relation $\tau\equiv R-\Lambda\approx 0$ in
UG. There are two distinctions, however. First, in the case of UG,
we have one completion function which involves one constant
parameter. In the case of spin-3, we have $d$ completion functions
involving $\displaystyle \frac{(d+1)(d+2)}{2}$ constant parameters.
Second, in the case of UG, the completion function does not depend
on space-time coordinates, while for $s=3$ there is explicit
$x$-dependence. We see that the number of modular parameters does
not directly correlate to the number of completion functions. Also,
completion functions can be explicitly $x$-dependent, even if the
Lagrangian is translation-invariant. Specific modular parameters
$\lambda,a,b,\omega$ are defined by asymptotics of the fields. If
the fields tend to zero at infinity, all the parameters vanish,
while the equation $\tau_\mu=0$ will remain a non-trivial relation
anyway.

Notice that field equations (\ref{EoMspin3}) admit the solutions
such that compatible with any modular parameters
$\lambda,a,b,\omega$ in the completion function (\ref{taus3}). Let
$\varkappa_{\mu\nu\rho}^{(0)}$ be a general solution vanishing at
infinity. It incudes the Cauchy data, corresponding to four local
physical DoF in $4d$ case. Double divergence of
$\varkappa_{\mu\nu\rho}^{(0)}$ inevitably vanishes. There is another
solution, $\varkappa_{\mu\nu\rho}$, with different asymptotics which
includes the same number of local Cauchy data and arbitrary modular
parameters:
\begin{equation}\label{kappa}
\begin{array}{c}
\displaystyle \varkappa_{\mu\nu\rho}=\varkappa_{\mu\nu\rho}^{(0)}+
A\Big[a_{(\mu}x_{\nu}x_{\rho)}-\frac{1}{d+2}\eta_{(\mu\nu}a_{\rho)}x_{\lambda}x^{\lambda}-\frac{2}{d+2}\eta_{(\mu\nu}x_{\rho)}a_{\lambda} x^{\lambda}\Big]\,+\\[3mm]
\displaystyle+\,B\Big[\eta_{(\mu\alpha}\omega^{\alpha\beta}x_{\beta}x_{\nu}x_{\rho)}-\frac{1}{d+2}\eta_{(\mu\nu}\eta_{\rho)\alpha}\omega^{\alpha\beta}x_{\beta}x_{\lambda}x^{\lambda}\Big]+
C\lambda\Big[x_{(\mu}x_{\nu}x_{\rho)}-\frac{3}{d+2}\eta_{(\mu\nu}x_{\rho)}x_{\lambda}x^{\lambda}\Big]\,+\\[3mm]
\displaystyle +\,Db^\lambda x_{\lambda}x_{(\mu}x_{\nu}x_{\rho)}+
Eb_{(\mu}x_{\nu}x_{\rho)}x_{\lambda}x^{\lambda}+F\eta_{(\mu\nu}b_{\rho)}x_{\lambda}x^{\lambda}x_{\sigma}x^{\sigma}+G\eta_{(\mu\nu}x_{\rho)}b^{\lambda}x_{\lambda}x_{\sigma}x^{\sigma}\,,
\end{array}
\end{equation}
where
\begin{equation}
\begin{array}{c}
\displaystyle A=\frac{3}{(d+4)(d-1)}\,, \quad B=\frac{6}{(d+4)(d+1)}\,, \quad C=\frac{1}{(d+4)(d-1)}\,,\\[3mm]
\displaystyle D=\frac{2(d^2+7d-6)}{(d+6)(d+4)(d^2-1)}\,, \quad
 E=-\frac{3(d^2+3d-6)}{(d+6)(d+4)(d^2-1)}\,,\\[3mm]
\displaystyle F=\frac{3(d^2+3d-6)}{(d+6)(d+4)(d+2)(d^2-1)}\,, \quad
G=-\frac{24d}{(d+6)(d+4)(d+2)(d^2-1)}\,.
\end{array}
\end{equation}
For the solution $\varkappa$, the double divergence of the field is a
general conformal Killing vector (\ref{Lambdas3}):
\begin{equation}\label{ddivkappa}
\partial^\nu\partial^\lambda \varkappa_{\mu\nu\lambda}=\Lambda_\mu(x;\lambda,
a,b,\omega)\, .
\end{equation}
Once we have the completion function $\tau_\mu$ (\ref{taus3}),
relation (\ref{Conseqspin3}) can be re-formulated as modified
Noether identity (\ref{GI}):
\begin{equation}\label{MNIspin3}
\displaystyle \partial^{\lambda}\frac{\delta S}{\delta
\varphi^{\lambda\mu\nu}} -\frac{d}{d+2} \Big(\partial_\mu\tau_\nu+
\partial_\nu\tau_\mu
-\frac{2}{d}\eta_{\mu\nu}\partial_\lambda\tau^\lambda\Big)\equiv 0\,.
\end{equation}
Given the modified Noether identity, the coefficient at the
equations defines unfree gauge variation of the field, while the
coefficient at completion function defines the equation constraining
the gauge parameters. In this way, we get unfree gauge symmetry of
Lagrangian (\ref{Lagrspin3}):
\begin{equation}\label{symspin3}
\displaystyle
\delta_\varepsilon\varphi_{\mu\nu\lambda}=\partial_{\mu}\varepsilon_{\nu\lambda}
+\partial_{\nu}\varepsilon_{\lambda\mu}+\partial_{\lambda}\varepsilon_{\mu\nu}\,,
\end{equation}
\begin{equation}
\label{transvs3}
\partial^\nu\varepsilon_{\nu\mu}=0 \, ,
\end{equation}
where the gauge parameters are symmetric traceless tensors
$\varepsilon_{\mu\nu}=\varepsilon_{\nu\mu},\,\,\varepsilon^\nu{}_\nu=0$.
\noindent
\subsection{Constrained Hamiltonian formalism  for $s=3$ case.}
\noindent Hamiltonian formalism for the theory (\ref{Lspins}) is
worked out in the article \cite{SKVORTSOV2008301}. Our analysis
extends the consideration of \cite{SKVORTSOV2008301} in two
respects. First, the article \cite{SKVORTSOV2008301} assumed that
fields vanish at infinity. We admit non-trivial boundary conditions
for the fields, and reveal contribution of the modular parameters to
the Hamiltonian constraints. Second, we demonstrate that involution
relations of constraints and Hamiltonian define the unfree gauge
symmetry.

We begin constructing the Hamiltonian formalism with $1+(d-1)$
decomposition of the fields such that accounts for the traceless
condition. The indices $\mu,\nu,\ldots=0,1,\ldots, d-1$ are split
into $0$ and $i,j,\ldots=1,\ldots, d-1$. Metrics
$\eta_{ij}=\text{diag}(-1,\ldots, -1)$. Introduce abbreviation
\begin{equation}\label{3d0}
    \varphi_{0ij}\equiv\tilde{\varphi}_{ij}+\frac{1}{d-1}\eta_{ij}\varphi\,,\qquad
    \eta^{ij}\tilde{\varphi}_{ij}=0,
\end{equation}
and notice the consequences of symmetry and traceless properties of
$\varphi_{\mu\nu\lambda}$:
\begin{equation}\label{exl0}
\displaystyle
\varphi^0{}_{00}=-\varphi^i{}_{i0}\equiv-\varphi\,, \qquad
\varphi^0{}_{0i}=-\varphi^j{}_{ji}\,;
\end{equation}
where $\varphi^{j}{}_{ji}=\eta^{jk}\varphi_{kji}$. Given relations
(\ref{3d0}), (\ref{exl0}),  Lagrangian (\ref{Lagrspin3}), being
expressed in terms of the variables $\varphi_{ijk},
\widetilde\varphi_{ij}, \varphi$, modulo total time derivative reads:
\begin{equation}\label{Ldecoms3}
\begin{array}{c}
\displaystyle \mathcal{L}=-\frac{1}{2}\dot{\varphi}_{ijk}\dot{\varphi}^{ijk}+\frac{3}{2}\dot{\varphi}_{ij}{}^j\dot{\varphi}^{ik}{}_k+\dot{\varphi}^2
+3\dot{\varphi}_{ijk}\partial^k\widetilde{\varphi}^{ji}-6\dot{\varphi}_{ij}{}^j\partial_k\widetilde{\varphi}^{ki}-\frac{3}{d-1}\dot{\varphi}_{ij}{}^j\partial^i\varphi+3\dot{\varphi}\partial_i\varphi^{ij}{}_j\,-
\\[3mm]
\displaystyle -\,\frac{1}{2}\Big(\partial_i\varphi_{jkl}\partial^i\varphi^{jkl}+3\partial_i\varphi_{jk}{}^k\partial^i\varphi^{jl}{}_l+3\partial_i\widetilde{\varphi}_{jk}\partial^i\widetilde{\varphi}^{jk}+\frac{d^2+d-8}{(d-1)^2}\partial_i\varphi\partial^i\varphi\Big)\,+
\\[3mm]
\displaystyle +\,\frac{3}{2}\Big(\partial^i\varphi_{ikl}\partial_j\varphi^{jkl}+\partial^i\varphi_{ik}{}^k\partial_j\varphi^{jl}{}_l+2\partial^i\widetilde{\varphi}_{ik}\partial_j\widetilde{\varphi}^{jk}+\frac{4}{d-1}\partial^i\widetilde{\varphi}_{ij}\partial^j\varphi\Big)\,.
\end{array}
\end{equation}
The Lagrangian does not include $\dot{\tilde\varphi}{}_{ij}$. Making
the Legendre transform w.r.t. $\dot\varphi{}_{ijk}$ and
$\dot\varphi$, the action is brought to the Hamiltonian form
\begin{equation}\label{SHs3}
\displaystyle S_{H}=\int d^dx \big(\Pi^{ijk}\dot{\varphi}_{ijk}+\Pi\dot{\varphi}-H-\widetilde{\varphi}^{ij}{\widetilde{T}}{}_{ij}\big)\,,
\end{equation}
where the Hamiltonian reads
\begin{equation}\label{Hs3}
\begin{array}{c}
\displaystyle H=-\frac{1}{2}\Pi^{ijk}\Pi_{ijk}+\frac{3}{2}\frac{1}{d}\Pi^{ij}{}_j\Pi_{ik}{}^k
+\frac{1}{4}\Pi^2+\frac{3}{d(d-1)}\Pi^{ij}{}_j\partial_i\varphi-\frac{3}{2}\Pi\partial_i\varphi^{ij}{}_j\,+
\\[3mm]
\displaystyle +\,\frac{1}{2}\Big(\partial_i\varphi_{jkl}\partial^i\varphi^{jkl}+3\partial_i\varphi_{jk}{}^k\partial^i\varphi^{jl}{}_l+\frac{d+3}{d}\partial_i\varphi\partial^i\varphi\Big)
-\frac{3}{2}\Big(\partial^i\varphi_{ikl}\partial_j\varphi^{jkl}-\frac{1}{2}\partial^i\varphi_{ik}{}^k\partial_j\varphi^{jl}{}_l\Big)\,,
\end{array}
\end{equation}
and
\begin{equation}\label{T1s3}
\displaystyle
{\widetilde{T}}_{ij}\equiv-3\Big(\partial^k\Pi_{kij}-\frac{1}{d-1}\eta_{ij}\partial_k\Pi^{kl}{}_l\Big)=0\,,
\qquad \eta^{ij}{\widetilde{T}}_{ij}\equiv 0\,,
\end{equation}
are the primary constraints, with $\widetilde{\varphi}^{ij}$ being
Lagrange multipliers.

Let us examine stability of primary constraints (\ref{T1s3}):
\begin{equation}\label{dotT1s3}
\displaystyle
\dot{{\widetilde{T}}}_{ij}=\{{\widetilde{T}}_{ij}\,,H_0\}=-\Big(\delta_{(i}^k\partial_{j)}
-\frac{1}{d-1}\eta_{ij}\partial^k\Big){T'}_k=0\,.
\end{equation}
where
\begin{equation}\label{Ti}
\displaystyle
{T'}_i=-3\big(\partial_i\Pi-\partial_i\partial^j\varphi_{jk}{}^k-2\Delta\varphi_{ij}{}^j+2\partial^j\partial^k\varphi_{kji}\big)\,.
\end{equation}
The coefficient at $T'{}_i$ in relation (\ref{dotT1s3}) is a linear
differential operator with the finite kernel. The equation for the
null-vectors of the operator reads
\begin{equation}\label{kers3-1-1}
 \partial_i\Lambda_j+   \partial_j\Lambda_i
 -\frac{1}{d-1}\eta_{ij}\partial_k\Lambda^k =0 \, .
\end{equation}
The equation above defines the conformal Killing vector field in
$(d-1)$-dimensional space. The space of conformal Killing vectors is
finite-dimensional. There is a subtlety, however. It concerns the
fact that the parameters defining the solution to equation
(\ref{kers3-1-1}) may be time-dependent. This can be understood from
the fact that solution of (\ref{kers3-1-1}) should explicitly depend
on space coordinates $x^i$, while the theory is Lorentz-invariant.
Then, Lorentz boost will inevitably bring time-dependence to any
solution of (\ref{kers3-1-1}). The time-dependence is fixed, as we
shall see below, by further stability conditions. Stability
condition (\ref{dotT1s3}) means that ${T'}_i$  (\ref{Ti}) reduce to
the solution of equation (\ref{kers3-1-1}),
 i.e. we arrive at secondary constraints
\begin{equation}\label{T2s3}
\displaystyle {T}_i\equiv {T'}_i-\Lambda_i(x)=0\,.
\end{equation}
Given the secondary constraints, they have to conserve. The
conservation condition reads:
\begin{equation}\label{dotT2s3}
\displaystyle
\dot{{T}}_i=\partial_0{T}_i+\{{T}_i\,,H\}
=-\,\partial_0\Lambda_i(x)-2\Big(\delta_i^{(j}\partial^{k)}-\frac{1}{d-1}\eta^{jk}\partial_i\Big){\widetilde{T}}_{jk}+\partial_i{T}
=0\,.
\end{equation}
Relation (\ref{dotT2s3}) means we have tertiary constraint
\begin{equation}\label{Ts3}
T\equiv T'+\Lambda_0(x)=0\,,
\end{equation}
where
\begin{equation}\label{T3}
\displaystyle
T'=-3\Big(\frac{1}{d-1}\partial^i\Pi_{ij}{}^{j}+\Delta\varphi\Big)\,,
\end{equation}
and $\Lambda_0(x)$ is connected with $\Lambda_i(x)$ of
(\ref{T2s3}) by the relation
\begin{equation}\label{kers3-1-2}
\displaystyle \partial_0\Lambda_i+\partial_i\Lambda_0=0\,.
\end{equation}
Given tertiary constraint (\ref{Ts3}), it has to conserve,
\begin{equation}\label{dotT3}
\displaystyle
\dot{{T}}=\partial_0{T}+\{{T}\,,H\}=\partial_0\Lambda_0-\frac{1}{d-1}\partial^i{T}_i-\frac{1}{d-1}\partial^i\Lambda_i=0\,.
\end{equation}
This relation does not result in any new constraint, while it is
consistent if $\Lambda_0(x)$ and $\Lambda_i(x)$ are connected by one more
relation
\begin{equation}\label{kers3-1-3}
\displaystyle \partial_0\Lambda_0-\frac{1}{d-1}\partial^i\Lambda_i=0\,.
\end{equation}
Relations (\ref{kers3-1-1}), (\ref{kers3-1-2}), (\ref{kers3-1-3})
taken together are just $1+(d-1)$ decomposition of conformal Killing
equations (\ref{Lambdas3}) in $d$ dimensions. So, $\Lambda_i(x)$,
$\Lambda_0(x)$ are the components of conformal Killing vector,
\begin{equation}\label{Lambdai}
\displaystyle \Lambda_i=a_i+2(\omega_{i0}x^0+\omega_{ij}x^j)+\lambda
x_i+2(b_0x^0+b_jx^j)x_i-b_i(x_0x^0+x_jx^j)\,,
\end{equation}
\begin{equation}\label{Lambda0}
\displaystyle \Lambda_0=a_0+2w_{0i}x^i+\lambda x_0+b_0(x_0x^0-x_ix^i)+2b_ix^ix_0\,.
\end{equation}
As soon as the Dirac-Bergrmann algorithm is completed, let us
summarize its results. Complete set of constraints reads:
\begin{equation}\label{Tcompls3}
\begin{array}{l}
\displaystyle \widetilde{T}{}_{ij}=-3\Big(\partial^k\Pi_{kij}-\frac{1}{d-1}\eta_{ij}\partial_k\Pi^{kl}{}_l\Big)\,,\\[3mm]
\displaystyle {T}_i=-3\big(\partial_i\Pi-\partial_i\partial^j\varphi_{jk}{}^k-2\Delta\varphi_{ij}{}^j+2\partial^j\partial^k\varphi_{kji}\big)-\Lambda_i\,,\\[3mm]
\displaystyle T=-3\Big(\frac{1}{d-1}\partial^i\Pi_{ij}{}^{j}+\Delta\varphi\Big)+\Lambda_0\,,
\end{array}
\end{equation}
where $\Lambda_i$, $\Lambda_0$ are defined by relations
(\ref{Lambdai}), (\ref{Lambda0}). All the constraints Poisson-commute to each other. There are non-trivial involution relations
between the constraints and Hamiltonian:
\begin{equation}\label{HT-s3}
\begin{array}{l}
\displaystyle \{\widetilde{T}_{ij}\,, H\}= -\Big(\delta_{(i}^k\partial_{j)}
-\frac{1}{d-1}\eta_{ij}\partial^k\Big){T}_k\,,\\[3mm]
\displaystyle \partial_0 T_i + \{{T}_{i}\,, H\}=-2\Big(\delta_i^{(j}\partial^{k)}-\frac{1}{d-1}\eta^{jk}\partial_i\Big){\widetilde{T}}_{jk}+\partial_i{T}
\,,\\[3mm]
\displaystyle \partial_0 T + \{T\,, H\}=-\frac{1}{d-1}\partial^i{T}_i\,.
\end{array}
\end{equation}
Once  all the constraints are known, and structure coefficients of
involution relations (\ref{T1gen}), (\ref{Tkgen}), (\ref{Tngen})
are identified, they define the unfree gauge variations of the
fields and Lagrange multipliers by the general rules (\ref{ugtqp}),
(\ref{ugtlambda}). Also the structure coefficients define the
equations (\ref{ugtc1}), (\ref{ugtc2}) imposed on the gauge
parameters. Given the constraints (\ref{Tcompls3}) and involution
relations (\ref{HT-s3}), we apply the general rules, and arrive at
unfree gauge symmetry transformations of the fields $\varphi_{ijk},
\varphi$ and Lagrange multipliers $\widetilde{\varphi}^{ij}$ :
\begin{equation}\label{s3ugt1}
\displaystyle
\delta_\varepsilon\varphi_{ijk}=3\partial_{(i}\widetilde{\varepsilon}_{jk)}+\frac{3}{d-1}\eta_{(ij}\partial_{k)}\varepsilon\,,
\qquad \delta_\varepsilon\varphi=3\partial_i\varepsilon^i \, ,
\end{equation}
\begin{equation}\label{s3ugt1}
\displaystyle \delta_\varepsilon
\widetilde{\varphi}^{ij}=\dot{\widetilde{\varepsilon}}{}^{ij}+\partial^i\varepsilon^j+\partial^j\varepsilon^{i}-\frac{2}{d-1}\eta^{ij}\partial_k\varepsilon^{k}\,.
\end{equation}
Upon substitution of structure coefficients of involution relations
(\ref{HT-s3}) into general relations (\ref{ugtc1}), (\ref{ugtc2}),
we get the constraints on gauge parameters for this model:
\begin{equation}\label{s3gpc1}
\displaystyle
\dot{\varepsilon}^{i}+\partial_j\widetilde{\varepsilon}^{ji}+\frac{1}{d-1}\partial^i\varepsilon=0\,,
\end{equation}
\begin{equation}\label{s3gpc2}
\displaystyle \dot{\varepsilon}-\partial_i\varepsilon^i=0\,.
\end{equation}
This unfree gauge symmetry is parameterized by $(d-1)$-tensors
$\widetilde{\varepsilon}^{ij}, \, \varepsilon^i, \, \varepsilon$.
Explicitly  covariant unfree gauge symmetry (\ref{symspin3}),
(\ref{transvs3}) of the original action (\ref{Lagrspin3}) is
parameterized  by  symmetric traceless tensor
$\varepsilon_{\mu\nu}$.  The gauge parameters
$\widetilde{\varepsilon}^{ij}, \, \varepsilon^i, \, \varepsilon$ of
Hamiltonian form of the symmetry can be viewed as $1+(d-1)$
decomposition of the  $d$-tensor parameter $\varepsilon_{\mu\nu}$:
\begin{equation}\label{parameterid}
\displaystyle
\varepsilon^0{}_0=-\varepsilon^i{}_i\equiv-\varepsilon\,, \qquad
\varepsilon_{0i}\equiv\varepsilon_i\,, \qquad
\varepsilon_{ij}\equiv\widetilde{\varepsilon}_{ij}+\frac{1}{d-1}\eta_{ij}\varepsilon\,.
\end{equation}
As we see in this example, the Hamiltonian algorithm of section 3
allows one to systematically identify all the unfree gauge symmetry
transformations and modular parameters of the model, though the
method is not explicitly covariant.

\subsection{BFV-BRST formalism  for $s=3$ case.}
\noindent In this subsection, we illustrate the general BFV-BRST
formalism of section 4 by the spin-3 model (\ref{SHs3}).

We begin with construction of the formalism by introducing the
ghosts of the minimal sector. The ghost pairs are assigned to every
constraint of the complete set (\ref{Tcompls3}):
\begin{equation}
\begin{array}{c}
\displaystyle \{{\widetilde{C}}{}^{ij}\,,{\widetilde{\bar{P}}}_{kl}\}=\frac{1}{2}\big(\delta^i_k\delta^j_l+\delta^i_l\delta^j_k\big)-\frac{1}{d-1}\eta^{ij}\eta_{kl}\,, \quad \{{C}{}^i\,,{\bar{P}}_i\}=\delta^i_j\,, \quad \{{C}\,,{\bar{P}}\}=1\,,\\[3mm]
\displaystyle \text{gh}\left({\widetilde{C}}{}^{ij}\right)=-\,\text{gh}\left({\widetilde{\bar{P}}}_{ij}\right)=\text{gh}\left({C}{}^i\right)=-\,\text{gh}\left({\bar{P}}_i\right)=\text{gh}\left({C}\right)=-\,\text{gh}\left({\bar{P}}\right)=1\,,\\[3mm]
\displaystyle \epsilon\left({\widetilde{C}}{}^{ij}\right)=\epsilon\left({\widetilde{\bar{P}}}_{ij}\right)=\epsilon\left({C}{}^i\right)=\epsilon\left({\bar{P}}_i\right)=\epsilon\left({C}\right)=\epsilon\left({\bar{P}}\right)=1\,.
\end{array}
\end{equation}
Hamiltonian BRST generator reads:
\begin{equation}\label{Qs3}
\displaystyle
Q_{\text{min}}={\widetilde{C}}{}^{ij}{\widetilde{T}}_{ij}+{C}{}^i{T}_i
+{C}{T}\,, \qquad
\{Q_{\text{min}}\,,Q_{\text{min}}\}=0\,.
\end{equation}
Given the Hamiltonian $H$ (\ref{Hs3}), and involution relations
(\ref{HT-s3}), the ghost-extended Hamiltonian reads:
\begin{equation}\label{Hscripts3}
\displaystyle
\mathcal{H}=H-{\widetilde{C}}{}^{ij}\partial_{(i}{\bar{P}}_{ j)}+{C}{}^i
\partial_i{\bar{P}}-{C}{}^i\partial^j{\widetilde{\bar{P}}}_{ji}-\frac{1}{d-1}{C}\partial^i{\bar{P}}_i\,,
\qquad \partial_0Q_{\text{min}}+\{Q_{\text{min}}\,,\mathcal{H}\}=0\, .
\end{equation}
According to the prescriptions of section 4, the non-minimal sector
ghosts are assigned only to the primary constraints:
\begin{equation}
\begin{array}{c}
\displaystyle \{{\widetilde{P}}{}^{ij}\,,{\widetilde{\bar{C}}}_{kl}\}=\frac{1}{2}\big(\delta^i_k\delta^j_l+\delta^i_l\delta^j_k\big)-\frac{1}{d-1}\eta^{ij}\eta_{kl}\,,\\[3mm]
\displaystyle \text{gh}\left({\widetilde{P}}{}^{ij}\right)=-\,\text{gh}\left({\widetilde{\bar{C}}}_{ij}\right)=1\,, \qquad \epsilon\left({\widetilde{P}}{}^{ij}\right)=\epsilon\left({\widetilde{\bar{C}}}_{ij}\right)=1\,.
\end{array}
\end{equation}
Also the momenta are introduced being canonically conjugate to the
Lagrange multipliers,
\begin{equation}
\begin{array}{c}
\displaystyle \{\widetilde{\varphi}{}^{ij}\,,\widetilde{\Pi}_{kl}\}=\frac{1}{2}\big(\delta^i_k\delta^j_l+\delta^i_l\delta^j_k\big)-\frac{1}{d-1}\eta^{ij}\eta_{kl}\,,\\[3mm]
\displaystyle \text{gh}\left(\widetilde{\varphi}{}^{ij}\right)=-\,\text{gh}\left(\widetilde{\Pi}_{ij}\right)=0\,, \qquad \epsilon\left(\widetilde{\varphi}{}^{ij}\right)=\epsilon\left(\widetilde{\Pi}_{ij}\right)=0\,.
\end{array}
\end{equation}
Complete Hamiltonian BRST generator reads:
\begin{equation}
\displaystyle
Q=Q_{\text{min}}+\widetilde{\Pi}_{ij}{\widetilde{P}}{}^{ij}\,.
\end{equation}
Lorentz-like gauge conditions should be imposed, being explicitly
resolved w.r.t. time derivatives of Lagrange multipliers. As
explained in the section 4, the number of gauges should be the same
as the number of primary constraints. So, we choose the following
gauges:
\begin{equation}
\displaystyle \dot{\widetilde{\varphi}}{}^{ij}-{\widetilde{\chi}}{}^{ij}=0\,, \qquad {\widetilde{\chi}}{}^{ij}\equiv-\big(\partial_k\varphi^{kij}-\frac{1}{d-1}\eta^{ij}\partial_k\varphi^{kl}{}_l\big)\,.
\end{equation}
Given the gauges, the gauge Fermion reads:
\begin{equation}
\displaystyle \Psi={\widetilde{\overline{C}}}_{ij}{\widetilde{\chi}}{}^{ij}+\widetilde{\varphi}{}^{ij}{\widetilde{\bar{P}}}_{ij}\,.
\end{equation}
Following the general rule (\ref{H-psi}), the gauge-fixed
Hamiltonian is constructed,
\begin{equation}
\begin{array}{c}
\displaystyle H_\Psi=\mathcal{H}+\{Q\,,\Psi\}=\mathcal{H}+{\widetilde{\varphi}}{}^{ij}{\widetilde{T}}_{ij}+{\widetilde{\Pi}}_{ij}{\widetilde{\chi}}{}^{ij}+{\widetilde{\bar{P}}}_{ij}{\widetilde{P}}{}^{ij}\,+\\[3mm]
\displaystyle+\,{\widetilde{\bar{C}}}_{ij}\{{\widetilde{\chi}}{}^{ij}\,,{\widetilde{T}}_{kl}\}{\widetilde{C}}{}^{kl}+{\widetilde{\bar{C}}}_{ij}\{{\widetilde{\chi}}{}^{ij}\,,{T}_k\}{C}{}^k+{\widetilde{\bar{C}}}_{ij}\{{\widetilde{\chi}}{}^{ij}\,,{T}\}{C}\,.
\end{array}
\end{equation}
As a result, we arrive at the gauge-fixed BRST-invariant Hamiltonian
action
\begin{equation}
\displaystyle S_{\text{BRST}}^\Psi=\int d^dx\Big[\Pi^{ijk}\dot{\varphi}_{ijk}+\Pi\dot{\varphi}+\widetilde{\Pi}_{ij}\dot{\widetilde{\varphi}}{}^{ij}+{\widetilde{\bar{P}}}_{ij}\dot{{\widetilde{C}}}{}^{ij}+{\bar{P}}_i\dot{{C}}{}^i+{\bar{P}}\dot{{C}}+{\widetilde{\bar{C}}}_{ij}\dot{{\widetilde{P}}}{}^{ij}-H_\Psi\Big]\,.
\end{equation}
Corresponding path integral reads:
\begin{equation}\label{Z-psi-s3}
\displaystyle Z_\Psi=\int [D\Phi]\exp\Big\{\frac{i}{\hbar}S_{\text{BRST}}^\Psi\Big\}\,, \qquad
\end{equation}
where
$\Phi=\big\{\Pi^{ijk}\,,\varphi_{ijk},\Pi\,,\varphi,\widetilde{\Pi}_{ij},\widetilde{\varphi}^{ij},{\widetilde{\bar{P}}}_{ij},{\widetilde{C}}{}^{ij},{\bar{P}}_i,{C}{}^i,{\bar{P}},
{C},{\widetilde{\bar{C}}}_{ij},{\widetilde{P}}{}^{ij}\big\}$.
\noindent Integration over momenta ${\widetilde{P}}{}^{ij},
{\widetilde{\bar{P}}}_{ij}$,  $\Pi_{ijk},\Pi$, leads to the
following result:
\begin{equation}\label{Zs3}
\displaystyle Z=\int [D\Phi']\exp\Big\{\frac{i}{\hbar}\int
d^dx\Big[\mathcal{L}+\widetilde{\Pi}_{ij}\partial_\mu{\varphi}{}^{\mu
ij}+\widetilde{\bar{C}}_{ij}\big(\square
\delta^i_\mu\delta^j_\nu+2\partial^i\delta^j_\mu\partial_\nu\big)C^{\mu\nu}
+{\bar{P}}_\mu\partial_\nu C^{\nu\mu}\Big]\Big\}\,,
\end{equation}
where $\Phi'=\big\{\varphi_{\mu\nu\rho},\widetilde{\Pi}_{ij},
{\widetilde{\bar{C}}}_{ij},C^{\mu\nu},{\bar{P}}_\mu\big\}$.
Ghosts $C, C^i, \widetilde{C}^{ij}$ can be viewed as $1+(d-1)$
decomposition of ghost $C^{\mu\nu}, C^\nu{}_\nu=0$ (\ref{C-constr}),
being $d$-dimensional symmetric  traceless tensor,
\begin{equation}
\displaystyle C^0{}_0=-C^i{}_i\equiv-\,C\,, \qquad
C_{0i}\equiv C_i\,, \qquad
C_{ij}\equiv\widetilde{C}_{ij}+\frac{1}{d-1}\eta_{ij}C\,.
\end{equation}
Expression $\partial_\nu C^{\nu\mu}$ in the end of exponential of
(\ref{Zs3}) can be viewed as a constraint imposed on the ghosts
which corresponds to the transversality condition imposed on gauge
parameters. The ghost momenta $\bar{P}_\mu\,:
\bar{P}_0\equiv-\bar{P}$, assigned to the secondary and
tertiary constraints, play the role  of Fourier multipliers at the
constraints imposed on ghosts for the original unfree gauge
symmetry. With this regard, relation (\ref{Zs3}) is seen to
reproduce the modified FP path integral (\ref{ZFP}), (\ref{FP-act})
for the original action (\ref{Lagrspin3}).

\section{Conclusion}
In this article we work out constrained Hamiltonian formalism
corresponding to the unfree gauge symmetry with gauge parameters
constrained by differential equations. In the Hamiltonian form, the
phenomenon of the unfree gauge symmetry has been clarified from
viewpoint of involution relations between Hamiltonian and
constraints. The key role is plaid by differential operators
$\Gamma$, being the coefficients in the involution relations
(\ref{T1gen}), (\ref{Tkgen}) such that stand at the constraints of
the next generation in the stability conditions of the previous
constraints. These structure coefficients define the unfree gauge
symmetry if they have a finite kernel. Even if $\Gamma$ are
non-degenerate (trivial kernel), but the inverse does not exist in
the class of differential operators, we have unfree gauge symmetry.
Given the structure coefficients of involution relations with these
properties, we arrive at the equations constraining the gauge
parameters (\ref{ugtc1}), (\ref{ugtc2}). In the best-known example
of the unfree gauge symmetry, the unimodular gravity, the kernel of
$\Gamma$ is one-dimensional, and the corresponding modular parameter
is the cosmological constant $\Lambda$. The modular parameters are
defined by the asymptotics of the fields. For example, even the free
spin-2 field theory \cite{Alvarez:2006uu}, \cite{Blas:2007pp} with
unfree gauge symmetry in Minkowski space (that corresponds to
linearized UG), admits solutions with non-vanishing $\Lambda$. These
solutions correspond to non-vanishing fields at infinity. Analogous
solutions with non-trivial modular parameters are noticed in section
5 for higher spin fields with unfree gauge symmetry. The dynamics
with non-trivial modular parameters are relevant upon inclusion of
interactions as we expect. This issue will be addressed elsewhere.
In section 4, we explain how the Hamiltonian BFV-BRST formalism is
adjusted for the case of unfree gauge symmetry. For the case when
there are no higher-order ghost vertices, we deduce from the
phase-space path integral the modified FP quantization rules such
that account for the unfree gauge symmetry by imposing corresponding
constraints on the ghosts. In this way, we see that the covariant
quantization rules for the systems with unfree gauge symmetry are
deduced from corresponding modification of Hamiltonian BFV-BRST
quantization.

\vspace{0.4 cm} \noindent \textbf{Acknowledgments.} The work is
supported by the Foundation for the Advancement of Theoretical
Physics and Mathematics ``BASIS".

\end{document}